\documentclass[11pt,a4paper,openright]{article}

\providecommand{\keywords}[1]{\textbf{\small{\textit{Index terms:}}} \small{#1}}

\usepackage[bookmarks,%
            a4paper,%
            breaklinks,%
            backref=false,%
            pdfhighlight=/I,%
            pdffitwindow=true,%
            pdfstartview=Fit,%
            pdfcenterwindow=true,%
            linkbordercolor={1 0 1}]%
            {hyperref}

\usepackage{amsmath}

\usepackage{booktabs}
\usepackage{graphicx}
\graphicspath{{expt/}}

\usepackage{times}
\usepackage{caption}
\usepackage{subcaption}

\usepackage{enumerate}

\usepackage[varg]{txfonts}
\usepackage{amsmath, amssymb}
\usepackage{verbatim}

\begin{document}
\title{Deterministic Models in Epidemiology: from Modeling to Implementation}
\author{Aresh Dadlani}
\date{March 18, 2013}
\maketitle
\begin{center}
\begin{tabular}{l r}
Co-authors: & Richard O. Afolabi, Hyoyoung Jung \\
Supervisors: & Prof. Khosrow Sohraby, Prof. Kiseon Kim
\end{tabular}
\end{center}

%To cite:
%\begin{quote}
%	 Aresh Dadlani, Richard O. Afolabi, Hyoyoung Jung, Khosrow Sohraby, and Kiseon Kim, ``Deterministic Models in Epidemiology: From Modeling to Implementation,'' Technical Report, Gwangju Institute of Science and Technology, Communications and Sensor Networks Laboratory, Gwangju, South Korea, March 2013.
%\end{quote}
\footnotetext{This work has been funded by the World Class University (WCU) Program at Gwangju Institute of Science and Technology (GIST) through a grant provided by the Ministry of Education, Science, and Technology (MEST) of South Korea (Project No. R31-10026).}
\footnotetext{Kindly cite as: ``Aresh Dadlani, Richard O. Afolabi, Hyoyoung Jung, Khosrow Sohraby, and Kiseon Kim, ``Deterministic Models in Epidemiology: From Modeling to Implementation,'' Technical Report, Gwangju Institute of Science and Technology, Communications and Sensor Networks Laboratory, Gwangju, South Korea, March 2013.'' }

\begin{abstract}
The abrupt outbreak and transmission of biological diseases has always been a long-time concern of humankind. For long, mathematical modeling has served as a simple and yet efficient tool to investigate, predict, and control spread of communicable diseases through individuals. A myriad of works on epidemic models and their variants have been reported in the literature. For better prediction of the dynamics of a particular disease, it is important to adopt the most suitable model. In this paper, we study some of the widely-appreciated deterministic epidemic models in which the population is divided into compartments based on the health status of each individual. In particular, we provide a demographic classification of such models and study each of them in terms of mathematical formulation, near equilibrium point stability properties, and disease outbreak threshold conditions (basic reproduction ratio). Furthermore, we discuss the various influential factors that need to be considered during epidemic modeling. The main objective of this article is to provide a basic understanding of the mathematical complexity incurred in deterministic epidemic models with the aid of graphical illustrations obtained through implementation.
\end{abstract}


\newpage
\pagenumbering{roman}
\tableofcontents

\listoftables
\listoffigures

\cleardoublepage
\setcounter{page}{1}
\pagenumbering{arabic}

\section{Introduction}
Throughout recorded history, human population has always been haunted by the emergence and re-emergence of infectious diseases. Several lives have been lost due to lack of knowledge on the dynamical behavior of epidemic outbreaks of contagious diseases and measures to confront them \cite{Dobson1996}, \cite{Oldstone1998}. For decades, scientists have toiled to understand the transmission characteristics of such diseases so as to devise control strategies to prevent further spread of infection. The field of science that studies such epidemic diseases and in particular, the factors that influence the incidence, distribution, and control of infectious diseases in human populations is called \emph{epidemiology}. In this regard, mathematical modeling has proven to serve useful in analyzing, predicting, evaluating, detecting, and implementing efficient control programs. Such analytical models accompanied by computer simulations serve as experimental tools for building, testing, and assessing theories and understanding the relationship between various parametric values involved.

Practical use of epidemic models depends on how closely they realize actual biological diseases in real world. To keep the models simple and tractable, many assumptions and relaxations are taken into consideration at each level of the process. However, even such simplified models often pose significant questions regarding the underlying mechanisms of infection spread and possible control approaches. Hence, adopting the apt epidemic model for prediction of real phenomenon is of great importance.

\begin{figure}[!t]
\centering
\includegraphics[width=5.0in]{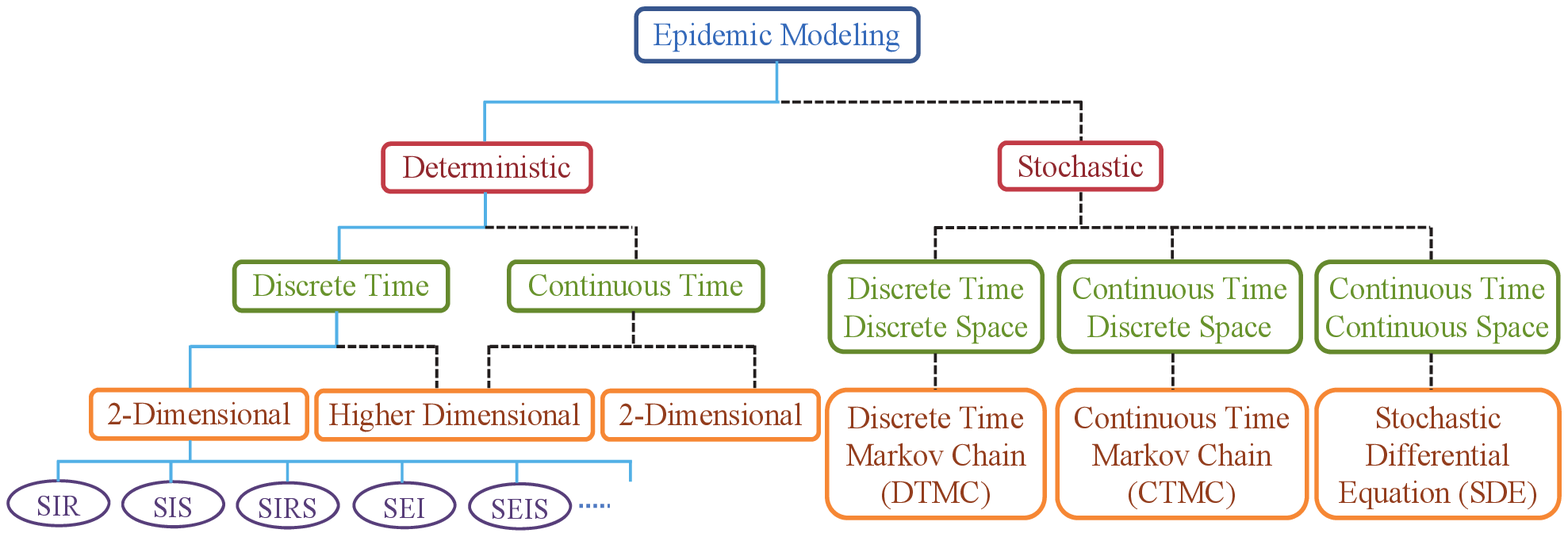}
\caption{Classification of various classes of epidemic models.}
\label{fig1_classification}
\end{figure}

Models that are useful in the study of infectious diseases at the population scale can be broadly classified into two types: \emph{deterministic} and \emph{stochastic}. Early models that were developed to study specific diseases such as tuberculosis, measles, and rubella were deterministic in nature. In deterministic models, the large population is divided into smaller groups called \emph{compartments} (or \emph{classes}) where each group represents a specific stage of the epidemic. Such models, often formulated in terms of a system of differential equations (in continuous time) or difference equations (in discrete time), attempt to explain what happens on the average at the population scale. A solution of a deterministic model is a function of time or space and is generally uniquely dependent on the initial data. On the other hand, a stochastic model is formulated in terms of a stochastic process which, in turn, is a set of random variables, $X(t;\omega)\equiv X(t)$, defined as $\{X(t;\omega)|t\in T and~ \omega\in \Omega\}$ where $T$ and $\Omega$ represent time and a common sample space, respectively.  The solution of a stochastic model is a probability distribution for each of the random variables. Such models capture the variability inherent due to demographic and environment variability and are useful under small population sizes. More specifically, they allow follow-up of each individual in the population on a chance basis \cite{Trottier2001}, \cite{Britton2010}. Discrete-time Markov chain (DTMC), continuous-time Markov chain (CTMC), and stochastic differential equation (SDE) models are three types of stochastic modeling processes which have been deeply covered in \cite{Allen2008}. Figure~\ref{fig1_classification} shows the different classes under which epidemic models have been studied in the literature. The connecting blue lines in the figure highlight the main scope of this report.

Needless to say, both deterministic and stochastic epidemic models have their own applications. Deterministic models are used to address questions such as: \textit{what fraction of individuals would be infected in an epidemic outbreak?}, \textit{what conditions should be satisfied to prevent and control an epidemic?}, \textit{what happens if individuals are mixed non-homogeneously?}, and so on \cite{Brauer2008}. While such models are preferable in studying a large population, stochastic epidemic models are useful for a small community and answer questions such as: \textit{how long is the disease likely to persist?}, \textit{what is the probability of a major outbreak?}, and the like \cite{Allen2008}. Hence, stochastic epidemic models are generalized forms of simple deterministic counterparts. However, unlike deterministic models, stochastic models can be laborious to set up and may need many simulation runs to yield useful predictions. They can become mathematically very complex and result in misperception of the dynamics \cite{Trottier2001}. To this end, we focus on some widely-used deterministic models which are relatively easier to conceive, set up, and implement using various computer softwares at disposal.

Deterministic epidemiology is believed to have started in the early twentieth century \cite{Hethcote2000}. In 1906, Hamer was the first to assume that the incidence (number of new cases per unit time) is proportional to the product of the number of susceptible and infective individuals in his model for measles epidemics \cite{Hamer1906}. The exponential growth in mathematical epidemiology was boosted by the acclaimed work of Kermack and McKendrick which was published in 1927 \cite{Kermack1927}. This paper laid out a foundation for modeling infections where all members of the population are assumed to be initially equally susceptible to the disease and confer complete immunity only after recovery. After decades of neglect, the Kermack-McKendrick model was brought back to prominence by Anderson \emph{et al.} in 1979 \cite{Anderson1979}. Since then several models have been developed addressing aspects such as passive and disease-acquired immunity, vaccination, quarantine, vertical transmission, disease vectors, age structure, social and sexual mixing groups, as well as chemotherapy \cite{Brauer2001, Chavez1989,Hethcote1994,Keeling2011}. Improved models have also been designed for diseases such as measles, chickenpox, smallpox, whooping cough, malaria, rabies, diphtheria, filariasis, herpes, syphilis, and HIV/AIDS \cite{Keeling2011}, \cite{Vynnycky2010}.

The main objective of this article is to help the reader gain an insight on the basics of deterministic compartmental modeling through implementation. In this work, we formulate some well-known models and derive their steady-state solutions. Since the models under study are non-linear in nature, we investigate their qualitative behavior near their corresponding equilibria using linearization method \cite{Khalil2001}. All models discussed in this paper have been implemented using \emph{Wolfram Mathematica} \cite{mathem}, the codes of which are freely available.

The remainder of this article is structured as follows. Section II provides a demographic classification of deterministic models along with notations and assumptions that will be used throughout the paper. In Section III, we present the classical epidemic model known as the \emph{susceptible}-\emph{infected}-\emph{recovered} (\emph{SIR}) model which forms the basis for the extended models that follow in Sections IV to VII, accompanied by their implementation results. In Section VIII, we present some additional factors that impact the behavior of epidemic models, followed by some conclusive remarks in Section IX.

% An example of a double column floating figure using two subfigures.
% (The subfigure.sty package must be loaded for this to work.)
% The subfigure \label commands are set within each subfigure command, the
% \label for the overall fgure must come after \caption.
% \hfil must be used as a separator to get equal spacing
%
%\begin{figure*}
%\centerline{\subfigure[Case I]{\includegraphics[width=2.5in]{subfigcase1.eps}
%\hfil
%\subfigure[Case II]{\includegraphics[width=2.5in]{subfigcase2.eps}
%\label{fig_second_case}}}
%\caption{Simulation results}
%\label{fig_sim}
%\end{figure*}

% An example of a floating table. Note that, for IEEE style tables, the
% \caption command should come BEFORE the table. Table text will default to
% \footnotesize as IEEE normally uses this smaller font for tables.
% The \label must come after \caption as always.
%
%\begin{table}
%% increase table row spacing, adjust to taste
%\renewcommand{\arraystretch}{1.3}
%\caption{An Example of a Table}
%\label{table_example}
%\centering
%% The array package and the MDW tools package offers better commands
%% for making tables than plain LaTeX2e's tabular which is used here.
%\begin{tabular}{|c||c|}
%\hline
%One & Two\\
%\hline
%Three & Four\\
%\hline
%\end{tabular}
%\end{table}

%~~~~~~~~~~~~~~~~~~~~~~~~~~~~~~~~~~~~~~~~~~~~~~~~~~~~~~~~~~~~~~~~~~~~~~~~~~~~~~~~~~~~~~~~~~~~~~~~~~~~~~~~~
%
%
%
%
%
%~~~~~~~~~~~~~~~~~~~~~~~~~~~~~~~~~~~~~ Demographic Classification ~~~~~~~~~~~~~~~~~~~~~~~~~~~~~~~~~~~~~~~~
\section{Demographic classification and notations}
In order to analyze the structure of epidemic models as well as the relation between their structure and the resulting dynamics, it is important to classify models as clear and simple as possible. In our study, we classify and study compartmental models based upon \emph{demography} or \emph{vital dynamics}. Demography relates to the study of characteristics of human populations such as birth, death, incidence of disease, and so on. Epidemic models with vital dynamics consider an open population with births and deaths while models without vital dynamics have a closed and fixed population with no demographic turnover.

For better realization, we assume that the \emph{law of mass action} holds for all models in this paper. This law states that if individuals in a population mix homogeneously, then the encounters between infected and susceptible individuals occur at a rate proportional to their respective numbers in the population \cite{Anderson1992}. In other words, the rate at which the susceptible population becomes infected is directly proportional to the product of the sizes of the two populations. Table~\ref{table_1_notation} summarizes the notations that will be used in model derivation henceforth. Note that $S$, $I$, $R$, and $E$ are used to represent the compartments in the epidemic model as well as the proportion of the corresponding compartments at any time instant $t$.
\begin{table}[!hb]
\renewcommand{\arraystretch}{1.2}
\caption{Notations used in model derivation}
\label{table_1_notation}
\centering
\begin{tabular}{|c||c|}
\hline
\bfseries Notation & \bfseries Definition\\
\hline\hline
$N$ & Total population size\\
$S$ or $S(t)$ & Number of \emph{susceptible} individuals at time $t$\\
$I$ or $I(t)$ & Number of \emph{infected} individuals at time $t$\\
$R$ or $R(t)$ & Number of \emph{recovered} (or \emph{removed}) individuals at time $t$\\
$E$ or $E(t)$ & Number of \emph{exposed} individuals at time $t$\\
$M$ or $M(t)$ & Number of \emph{passively immune infants} at time $t$\\
$\beta$ & Contact (or transmission) rate\\
$\gamma$ & Recovery rate\\
$1/\varepsilon$ & Average latent period\\
$\nu$ & Loss of immunity rate of recovered individuals\\
$b$ & Birth rate\\
$\mu$ & Death rate\\
$\mathcal{R}_0$ & Basic reproduction number (or ratio)\\
$e_i$ & Equilibrium point indexed at $i$\\
$X^*$ & Equilibrium value of class $X$; $X \in \{S,I,R,E\}$\\
$DFE$ & Disease-free Equilibrium\\
$EE$ & Endemic Equilibrium\\
%$J$ & Jacobian matrix\\
$\lambda_i$ & Eigenvalue indexed at $i$\\ 
\hline
\end{tabular}
\end{table}

%~~~~~~~~~~~~~~~~~~~~~~~~~~~~~~~~~~~~~~~~~~~~~~~~~~~~~~~~~~~~~~~~~~~~~~~~~~~~~~~~~~~~~~~~~~~~~~~~~~~~~~~~~
%
%
%
%
%
%~~~~~~~~~~~~~~~~~~~~~~~~~~~~~~~~~~~~~~~~~~~~~ SIR Model ~~~~~~~~~~~~~~~~~~~~~~~~~~~~~~~~~~~~~~~~~~~~~~~~~
\section{The basic \emph{SIR} model}
\begin{figure}[!t]
\centering
\includegraphics[width=2.0in]{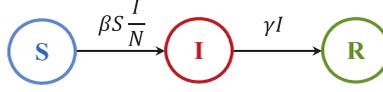}
\caption{Basic $SIR$ model without vital dynamics.}
\label{fig2_SIR}
\end{figure}

In their first paper, Kermack and McKendrick created a model in which the population is divided into three compartments: \emph{susceptible} ($S$), \emph{infected} ($I$), and \emph{recovered} ($R$) \cite{Kermack1927} as illustrated in Figure~\ref{fig2_SIR}. They assumed that all individuals are mutually equally susceptible to the disease and that complete immunity is obtained merely after recovery from infection. Moreover, they also assumed that the duration of the disease is same as the duration of infection with constant transmission and recovery rates. Based upon the demographic classification, the epidemic and endemic $SIR$ models are studied below.

\subsection{\emph{SIR} model without vital dynamics}
For a closed population of size $N$, we assume that the mixing of individuals is homogeneous and the law of mass action holds. Also, for large classes of communicable diseases, it is more realistic to consider a \emph{force of infection} that depends on the fraction of infected population with respect to the total constant population $N$, rather than the absolute number of infectious subjects. Based upon this assumption, the \emph{standard} disease incidence rate is defined as $\beta S I/N$ and the overall rate of recovery is given as $\gamma I$. In spite of the above simplifying assumptions, the resulting non-linear system does not admit a closed-form solution. Nevertheless, we shall see how significant results can be derived analytically. Figure~\ref{fig2_SIR} can be translated into the following set of differential equations:
\begin{eqnarray}
\label{eqn123_SIR_wvd}
\frac{dS}{dt}&{}={}&-\beta S \frac{I}{N},\\		%(1)
\frac{dI}{dt}&{}={}&\beta S \frac{I}{N} -\gamma I,\\		%(2)
\frac{dR}{dt}&{}={}&\gamma I.		%(3)
\end{eqnarray}

Summing up (1), (2), and (3) yields zero which implies that the population is of constant size with $S+I+R=N$. Dividing (2) by (1) gives:
\begin{equation}
\label{eqn4_SIR_wvd}
\frac{dI}{dS}=\frac{\gamma N}{\beta S}-1.		%(4)
\end{equation}

Assume that the population is susceptible up to time zero at which a relatively small number, $I(0)$, become infected. Thus, at $t=0$, $S(0)=N-I(0)$ and $R(0)=0$. As time approaches infinity, $\lim_{t \rightarrow \infty} I(t)=0$, $\lim_{t \rightarrow \infty} S(t)=S(\infty)$, and the number of individuals that have been infected is $S(0)-S(\infty)$. Integrating (4) leads to:
\begin{equation}
\label{eqn5_SIR_wvd}
I(\infty)-I(0)=\frac{\gamma N}{\beta} \ln \left(\frac{S(\infty)}{S(0)}\right)-S(\infty)+S(0)+c,	%(5)
\end{equation}
where $c$ is constant and $S(\infty)$ is the proportion of susceptibles at the end of the epidemic. Since the initial infection is small, (5) further reduces to:
\begin{equation}
\label{eqn6_SIR_wvd}
\ln \frac{S(\infty)}{S(0)}=\frac{\beta}{\gamma} \left(\frac{S(\infty)}{N}-1\right)+c^{'},	%(6)
\end{equation}
where $c^{'}$ denotes some constant. Defining $\mathcal{R}_0=\beta/\gamma$ as the \emph{basic reproduction ratio}, we see in Figure~\ref{fig3_SIR_wvd_Densities} that for $\mathcal{R}_0>1$, a small infection to the population would create an epidemic. $\mathcal{R}_0$ describes the total number of secondary infections produced when one infected individual is introduced into a disease-free population. The importance of the role of $\mathcal{R}_0$ can be seen by rewriting (2) as follows:
\begin{figure*}[!t]
        \centering
        \begin{subfigure}[b]{0.485\textwidth}
                \centering
                \includegraphics[width=\textwidth]{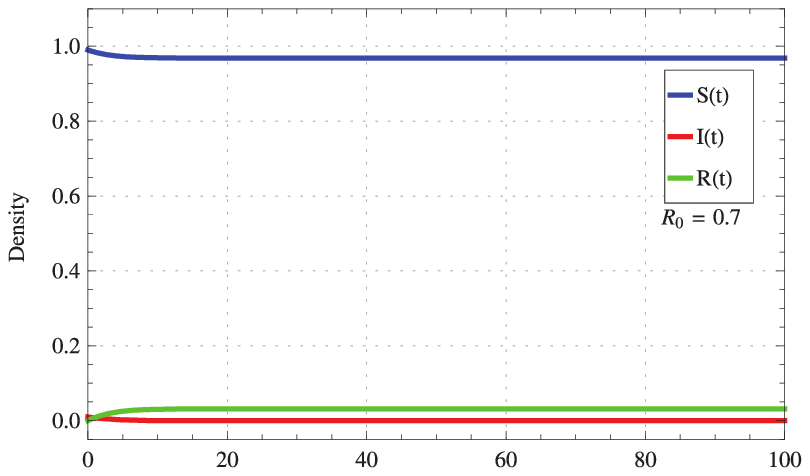}
                \caption{$\mathcal{R}_0=0.7$, $\beta=0.7$, and $\gamma=1$.}
                \label{fig3_1}
        \end{subfigure}%
        ~
        \begin{subfigure}[b]{0.485\textwidth}
                \centering
                \includegraphics[width=\textwidth]{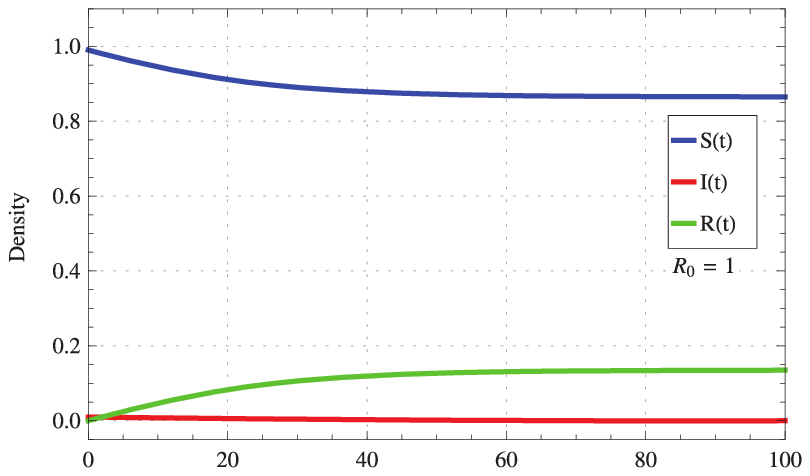}
                \caption{$\mathcal{R}_0=1$, $\beta=0.5$, and $\gamma=0.5$.}
                \label{fig3_2}
        \end{subfigure}
        
        ~
        
        \begin{subfigure}[b]{0.485\textwidth}
                \centering
                \includegraphics[width=\textwidth]{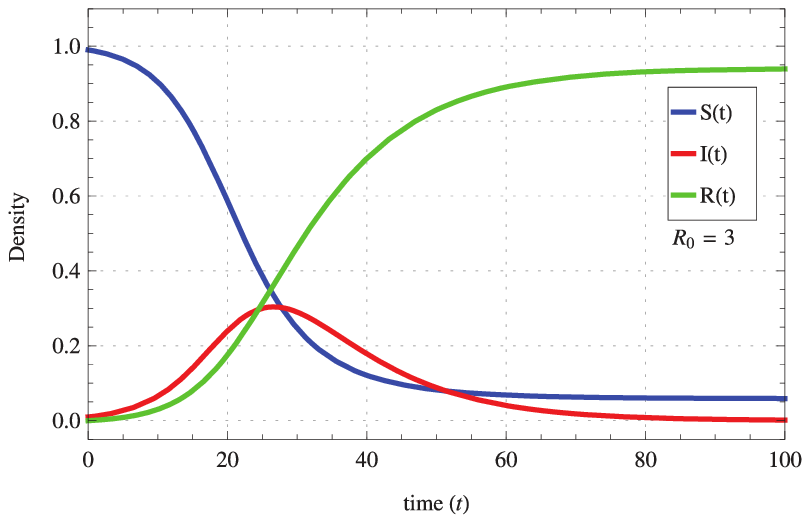}
                \caption{$\mathcal{R}_0=3$, $\beta=0.3$, and $\gamma=0.1$.}
                \label{fig3_3}
        \end{subfigure}
        ~
        \begin{subfigure}[b]{0.485\textwidth}
                \centering
                \includegraphics[width=\textwidth]{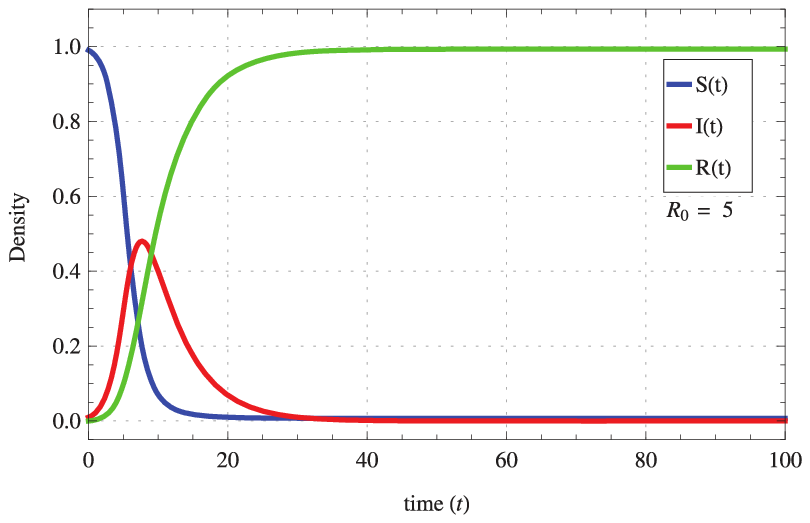}
                \caption{$\mathcal{R}_0=1$, $\beta=1.0$, and $\gamma=0.2$.}
                \label{fig3_4}
        \end{subfigure}
        \caption[Density versus time for $SIR$ model without vital dynamics]{Density versus time for $SIR$ model without vital dynamics where $N=1$, $S(0)=0.99$, $I(0)=0.01$, and $R(0)=0$.}
        \label{fig3_SIR_wvd_Densities}
\end{figure*}

\begin{equation}
\label{eqn7_SIR_wvd}
\frac{dI}{dt}=(\mathcal{R}_0 \frac{S}{N}-1) \gamma I.		%(7)
\end{equation}

In order to avoid an epidemic, (7) should be non-positive. This is possible only if $\mathcal{R}_0 S(0) \leq N$. On the other hand, if $\mathcal{R}_0 S(0) > N$, then (7) is positive and thus, there will be an epidemic outbreak. Figure~\ref{fig3_SIR_wvd_Densities} illustrates the limiting values of the $S$, $I$, and $R$ compartments for different values of $\mathcal{R}_0$ where $N$ is normalized to 1.

\subsection{\emph{SIR} model with vital dynamics}
Inclusion of demographic dynamics may permit a disease to persist in a population in the long term. A disease is said to be \emph{endemic} if it remains in a population for over a decade or two. Due to the long time period involved, an endemic disease model must include births as a source of new susceptibles and natural deaths in each compartment. In our study of the endemic $SIR$ model, we consider constant birth and death rates. Using the notations in Table~\ref{table_1_notation}, the scheme in Figure~\ref{fig4_SIR} can be expressed mathematically as:
\begin{figure}[!t]
\centering
\includegraphics[width=2.3in]{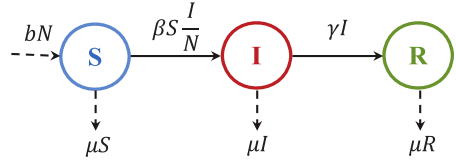}
\caption{Basic $SIR$ model with vital dynamics.}
\label{fig4_SIR}
\end{figure}
\setlength{\arraycolsep}{0.0em}
\begin{eqnarray}
\label{eqn8910_SIR_vd}
\frac{dS}{dt}&{}={}&b N - \beta S \frac{I}{N} - \mu S,\\		%(8)
\frac{dI}{dt}&{}={}& \beta S \frac{I}{N} -(\gamma+\mu) I,\\		%(9)
\frac{dR}{dt}&{}={}&\gamma I - \mu R.							%(10)
\end{eqnarray}

Assuming that $b$ equals $\mu$, we can easily see that the sum of the above three equations yields zero when $S+I+R=N$ holds in a non-varying population. Moreover, we observe that the average time of an infection is $1/(\gamma+\mu)$, and since the infectious individuals infect others at rate $\beta$, $\mathcal{R}_0$ is defined as $\beta/(\gamma+\mu)$.

\subsubsection{Existence of equilibria}
\begin{figure}[!t]
        \centering
        \begin{subfigure}[b]{0.485\textwidth}
                \centering
                \includegraphics[width=\textwidth]{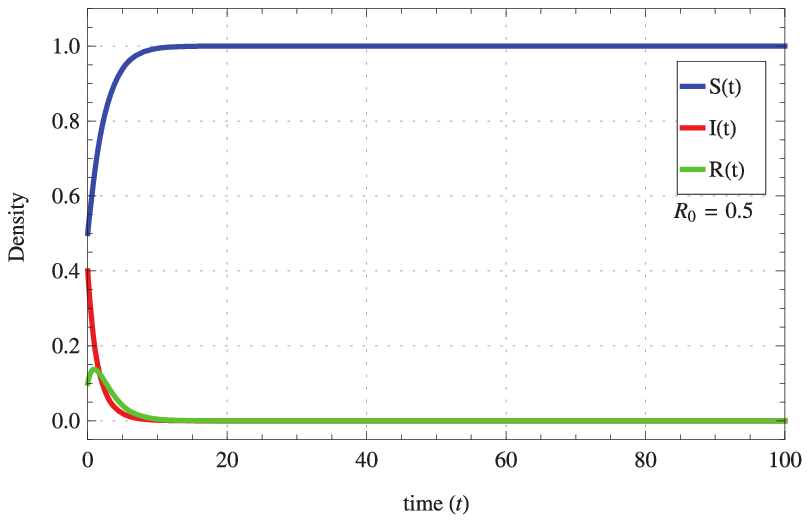}
                \caption{$\mathcal{R}_0=0.5$, $\beta= .5$, $\gamma=0.4$, and $\mu=0.6$.}
                \label{Fig5a}
        \end{subfigure}      
        ~
        \begin{subfigure}[b]{0.485\textwidth}
                \centering
                \includegraphics[width=\textwidth]{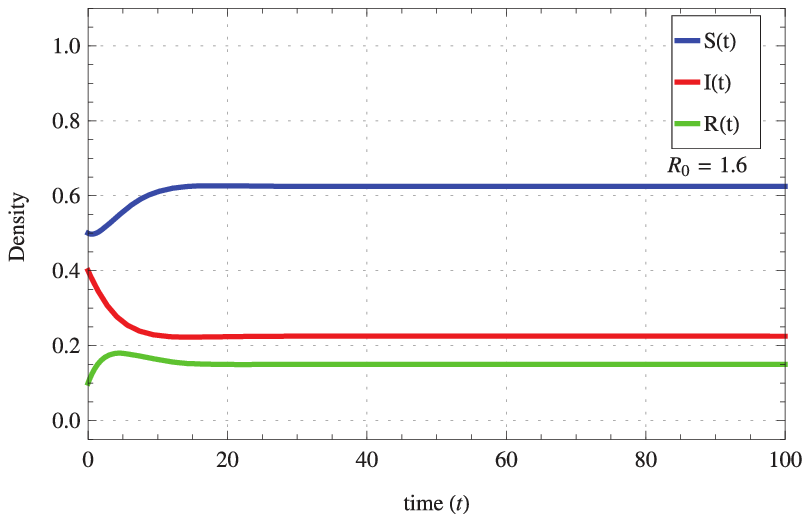}
                \caption{$\mathcal{R}_0=1.6$, $\beta=0.8$, $\gamma=0.2$, and $\mu=0.3$.}
                \label{Fig5b}
        \end{subfigure}
        \caption[Density versus time for $SIR$ model with vital dynamics]{Density versus time for $SIR$ model with vital dynamics where $N=1$, $S(0)=0.5$, $I(0)=0.4$, and $R(0)=0.1$.}
        \label{fig5_SIR_vd_Densities}
\end{figure}

By setting the left-hand side of the (8)-(10) to zero and solving for $S$, $I$, and $R$, we obtain the following two steady states (or equilibrium points) \cite{Khalil2001}:
\begin{equation}
\label{eqn11_SIR_vd}
  \begin{split}
    e_1:(S^*,I^*,R^*)&=(N, 0, 0),\\		%(11)
e_2:(S^*,I^*,R^*)&=\left(\frac{N}{\mathcal{R}_0}, N c_1 (\mathcal{R}_0-1), N c_2 (\mathcal{R}_0-1)\right), 
  \end{split}
\end{equation}
where $c_1=\mu/\beta$ and $c_2=\gamma/\beta$. Points $e_1$ and $e_2$ denote the \emph{disease-free equilibrium} ($DFE$) and \emph{endemic equilibrium} ($EE$) points, respectively. Figure~\ref{Fig5a} depicts the system in a disease-free steady state when $\mathcal{R}_0\leq 1$, whereas Figure~\ref{Fig5b} shows the occurrence of an endemic as the infected population reaches a limiting value of 0.225 when $\mathcal{R}_0 >1$. The stability of the system is driven by $\mathcal{R}_0$ as it can be observed by linearizing the system of equations at these points.

\subsubsection{Equilibria stability analysis}
The local stability of the model at these equilibrium points is analyzed via linearization. The Jacobian matrix for (8) and (9) is given as:
\renewcommand{\arraystretch}{2.0}
\begin{equation}
\label{eqn12_SIR_vd}
J=\begin{bmatrix}
  -\mu-\beta \dfrac{I}{N}  &\quad -\beta \dfrac{S}{N}\\
         \beta \dfrac{I}{N}   &\quad \beta \dfrac{S}{N}-(\gamma+\mu)		%(12)
\end{bmatrix}.
\end{equation}

Evaluating the above matrix at $e_1$ and solving the characteristic equation, $det(J-\mathbf{\lambda I})=0$, where $\mathbf{I}$ is the identity matrix of size 2, results in the following pair of eigenvalues:
\begin{eqnarray}
\label{eqn13_SIR_vd}
(\lambda_1,\lambda_2)|_{e_1}&{}={}&(-\mu, \beta-\gamma-\mu).		%(13)
\end{eqnarray}

In order to be a \emph{stable node}, both eigenvalues should be negative. Therefore, $e_1$ is stable when $\beta < \gamma+\mu$ (or equivalently $\mathcal{R}_0 < 1$) and unstable when $\beta > \gamma+\mu$. Similarly, the Jacobian matrix evaluated at $e_2$ is as below:
\renewcommand{\arraystretch}{2.0}
\begin{equation}
\label{eqn14_SIR}
J=\begin{bmatrix}
  -\mu \mathcal{R}_0 &\quad -\gamma-\mu\\
         \mu (\mathcal{R}_0-1) &\quad 0			%(14)
\end{bmatrix}.
\end{equation}

One can easily observe that the determinant of (14) is positive as long as $\mathcal{R}_0>1$. Hence, the endemic equilibrium is stable only when $\mathcal{R}_0>1$.
\setlength{\textfloatsep}{20pt}
\begin{figure*}[!t]
        \centering
        \begin{subfigure}[b]{0.48\textwidth}
                \centering
                \includegraphics[width=\textwidth]{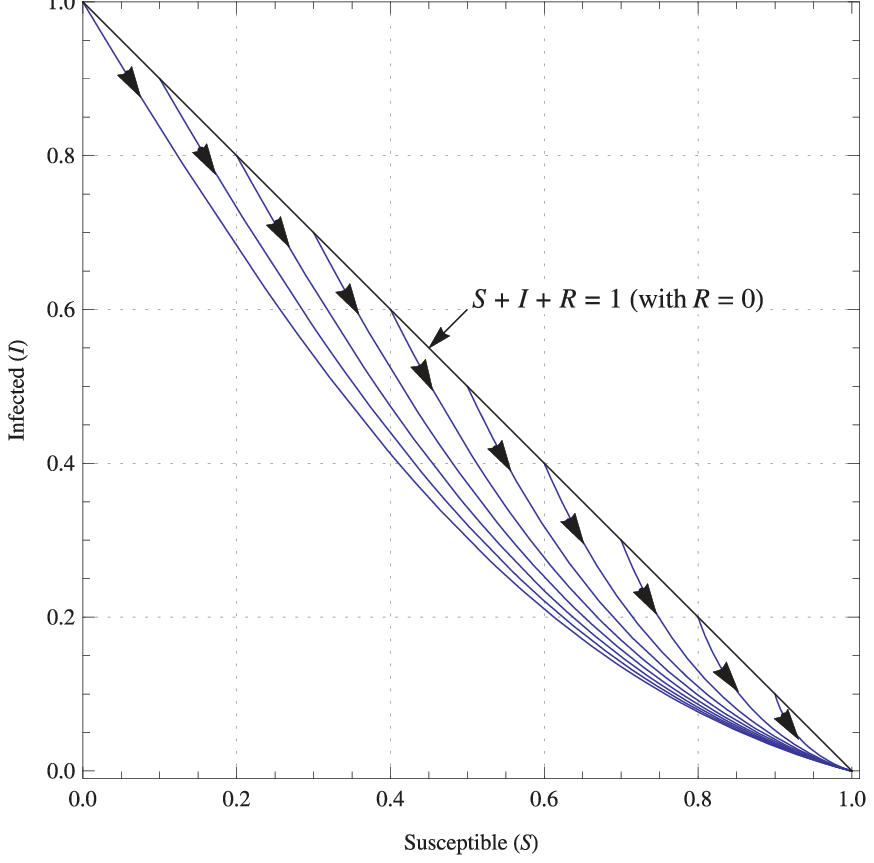}
                \caption{Disease-free equilibrium $(e_1)$.}
                \label{Fig6a}
        \end{subfigure}
        ~  
        \begin{subfigure}[b]{0.48\textwidth}
                \centering
                \includegraphics[width=\textwidth]{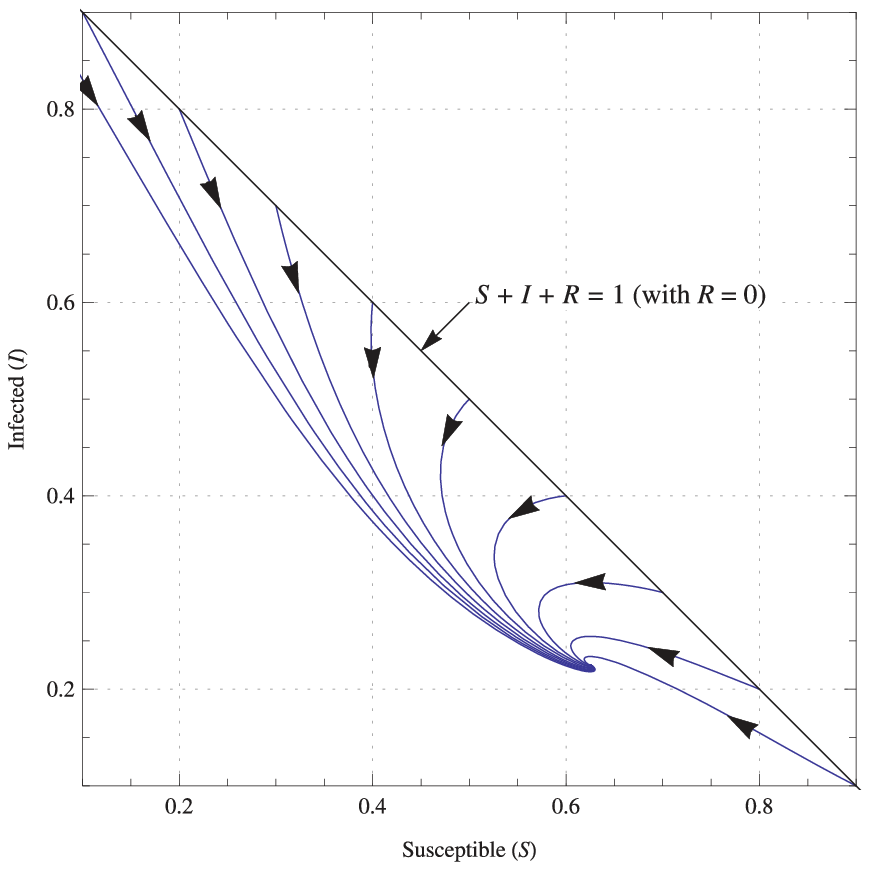}
                \caption{Endemic equilibrium $(e_2)$.}
                \label{Fig6b}
        \end{subfigure}
        \caption[Phase portrait of $SIR$ model with vital dynamics]{Phase portrait of $SIR$ model with vital dynamics for (a) $\mathcal{R}_0 \leq 1$ where the system converges to $e_1=(1,0)$ and for (b) $\mathcal{R}_0 > 1$ where it converges to $e_2=(0.625,0.225)$.}
        \label{fig6_SIR_portrait}
\end{figure*}

The \emph{S-I phase portrait} in Figure~\ref{fig6_SIR_portrait} shows how the model approaches the $DFE$ and $EE$ points with different initial values for $S(0)$ and $I(0)$. For the sake of simplicity, $N$ has been normalized to 1. As depicted in Figure~\ref{Fig6a}, for $\mathcal{R}_0 = 0.5$, the system eventually ends up at $(1,0)$, irrespective of the initial values of $S(0)$ and $R(0)$. On the other hand, an endemic occurs at $(0.625, 0.225)$ for $\mathcal{R}_0=1.6$ as in Figure~\ref{Fig6b}.

\begin{figure}[!t]
\centering
\includegraphics[width=3.2in]{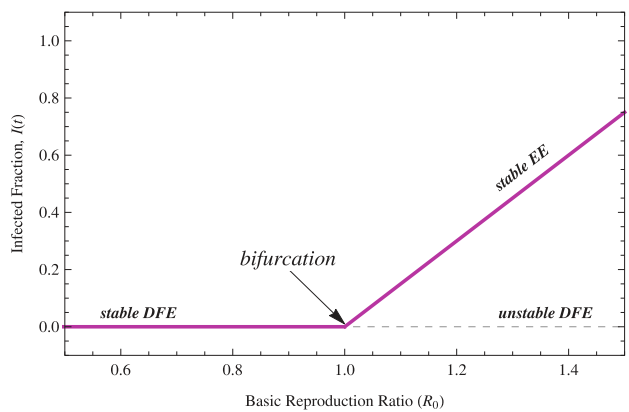}
\caption[Bifurcation diagram for the endemic $SIR$ model]{Bifurcation diagram for the endemic $SIR$ model where the equilibrium points change stability properties at $\mathcal{R}_0=1$.}
\label{fig7_SIR_Bifurcation}
\end{figure}

The phenomenon in which a parameter variation causes the stability of an equilibrium to change is known as \emph{bifurcation} \cite{Khalil2001}. In continuous systems, this corresponds to the real part of an eigenvalue of an equilibrium passing through zero. With the basic reproduction number as the \emph{bifurcation parameter}, Figure~\ref{fig7_SIR_Bifurcation} shows a \emph{transcritical} bifurcation where the equilibrium points persist through the bifurcation, but their stability properties change. Hence, we can conclude that:
\begin{equation}
\label{eqn15_SIR_vd}
  \begin{split}
    DFE&:\mathcal{R}_0 \leq 1 \Rightarrow \lim_{t \to \infty}(S(t), I(t), R(t)) = e_1,\\
 EE&:\mathcal{R}_0 > 1 \Rightarrow \lim_{t \to \infty}(S(t), I(t), R(t)) = e_2.			%(15)
  \end{split}
\end{equation}

A few recent studies have revealed interesting bifurcation behaviors in $SIR$ models incorporated with factors such as  varying immunity period, saturated treatment, and vaccination. We refer the interested reader to \cite{Onofrio2007,Jiang2009,Blyuss2010,Wang2012} and the references therein for more details.

%~~~~~~~~~~~~~~~~~~~~~~~~~~~~~~~~~~~~~~~~~~~~~~~~~~~~~~~~~~~~~~~~~~~~~~~~~~~~~~~~~~~~~~~~~~~~~~~~~~~~~~~~~
%
%
%
%
%
%~~~~~~~~~~~~~~~~~~~~~~~~~~~~~~~~~~~~~~~~~~~~~ SIS Model ~~~~~~~~~~~~~~~~~~~~~~~~~~~~~~~~~~~~~~~~~~~~~~~~~
\section{The $SIS$ model}
\begin{figure}[!t]
\centering
\includegraphics[width=1.3in]{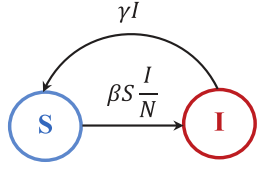}
\caption{The $SIS$ model without vital dynamics.}
\label{fig8_SIS_vd}
\end{figure}

For viral diseases, such as measles and chickenpox, where the recovered individuals, in general, gain immunity against the virus, the $SIR$ model is applicable. However, there exist certain bacterial diseases such as gonorrhoea and encephalitis that do not confer immunity. In such diseases, an infectious individual is allowed to recover from the infection and return unprotected to the susceptible class where he/she is prone to get infected again. Cases as such can be modeled using the \emph{susceptible-infected-susceptible} (\emph{SIS}) model as shown in Figure~\ref{fig8_SIS_vd}, where the model variables are defined in Table~\ref{table_1_notation}.

\subsection{SIS Model without vital dynamics}
In a fixed population, where there is no birth or death and individuals recover from the disease at the \emph{per capita} rate of $\gamma$, the simplest form of the model in Figure~\ref{fig8_SIS_vd} is given by:
\begin{eqnarray}
\label{eqn1617_SIS_wvd}
\frac{dS}{dt}&{}={}&\gamma I-\beta S \frac{I}{N},\\		%(16)
\frac{dI}{dt}&{}={}& \beta S \frac{I}{N}-\gamma I.		%(17)
\end{eqnarray}

By substituting $S=N-I$ in (17), the system above can be reduced to:
\begin{eqnarray}
\label{eqn18_SIS_wvd}
\frac{dI}{dt}&{}={}& (\beta-\gamma) I - \frac{\beta}{N} I^2.		%(18)
\end{eqnarray}

Solving (\ref{eqn18_SIS_wvd}) analytically with $I(0)=I_0$ gives the solution for the complete system at time $t$ as follows:
\begin{eqnarray}
\label{eqn1920_SIS_wvd}
S(t)&{}={}&N-I(t),\\		%(19)
I(t)&{}={}&\frac{(\beta - \gamma) N I_0}{(\beta - \gamma) N e^{-(\beta - \gamma) t}+ \beta I_0 \left[1-e^{-(\beta - \gamma) t}\right]} .		%(20)
\end{eqnarray}

The behavior of the system in long-term can be inferred by looking at the possible values of $(\beta-\gamma)$ that make (20) feasible. If $(\beta-\gamma) > 0$, then $e^{-(\beta-\gamma)t} \to 0$ as $t \to \infty$. This can be written as:
\begin{eqnarray}		%(21)
\lim_{t \rightarrow \infty} I(t)=\frac{(\beta-\gamma) N I_0}{\beta I_0}=\left(1-\frac{\gamma}{\beta}\right) N.
\end{eqnarray}

If $(\beta-\gamma)<0$, then $e^{-(\beta-\gamma)t} \to \infty$ as $t \to \infty$ and thus, $lim_{t \rightarrow \infty} I(t)=0$.

\subsubsection{Existence of equilibria}
There exists two equilibrium points for this model which can be obtained by setting $\frac{dI}{dt}=0$ in (18) and solving for $S$ and $I$. With $\mathcal{R}_0$ as $\beta/\gamma$, we get:
\begin{figure}[!t]
        \centering
        \begin{subfigure}[b]{0.485\textwidth}
                \centering
                \includegraphics[width=\textwidth]{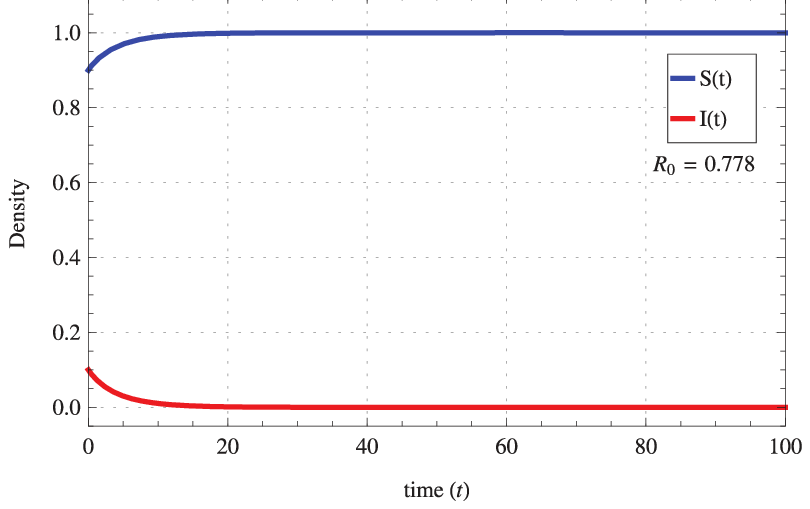}
                \caption{$\mathcal{R}_0=0.78$, $\beta=0.7$, and $\gamma=0.9$.}
                \label{Fig9a}
        \end{subfigure}
        ~
        \begin{subfigure}[b]{0.485\textwidth}
                \centering
                \includegraphics[width=\textwidth]{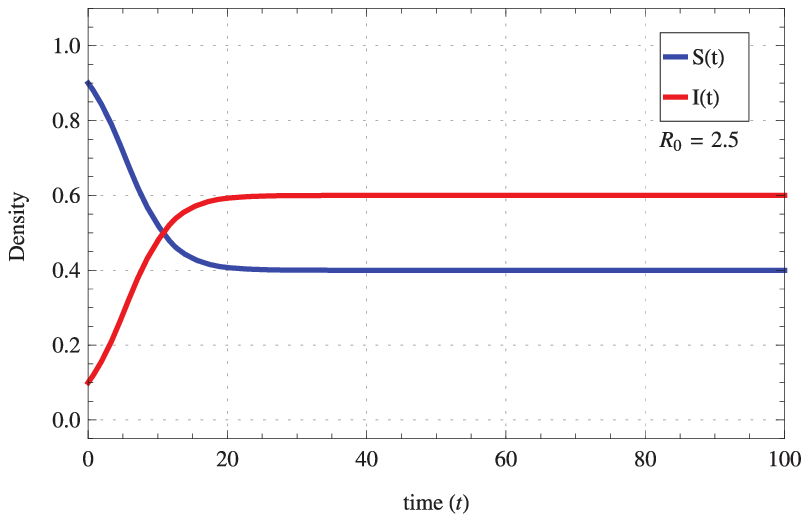}
                \caption{$\mathcal{R}_0=2.5$, $\beta=0.5$, and $\gamma=0.2$.}
                \label{Fig9b}
        \end{subfigure}
        \caption[Density versus time for $SIS$ model without vital dynamics]{Density versus time for $SIS$ model without vital dynamics where $N=1$, $S(0)=0.9$, and $I(0)=0.1$.}
        \label{fig9_Densities}
\end{figure}
\begin{equation}
\label{eqn22_SIR_vd}
  \begin{split}
    e_1:(S^*,I^*)&=(N, 0),\\			%(22)
e_2:(S^*,I^*)&=\left(\frac{N}{\mathcal{R}_0}, \frac{N}{\mathcal{R}_0}(\mathcal{R}_0-1)\right),
  \end{split}
\end{equation}
where $e_1$ and $e_2$ denote the $DFE$ and $EE$ points, respectively. In terms of the basic reproduction number, if $\mathcal{R}_0 \leq 1$, the pathogen dies out as illustrated in Figure~\ref{Fig9a} because the infection in one individual cannot replace itself. If $\mathcal{R}_0 > 1$, an existing infectious individual leads to more than one infection thus, spreading the pathogen in the population as seen in Figure~\ref{Fig9b}.

\subsubsection{Equilibria stability analysis}
The Jacobian matrix constructed from (16) and (17) is as follows:
\begin{equation}
\label{eqn14_SIR_vd}
J=\begin{bmatrix}
		-\beta \dfrac{I}{N}  &\quad \gamma - \beta \dfrac{S}{N}\\		%(23)
		\beta \dfrac{I}{N}   &\quad \beta \dfrac{S}{N} - \gamma 
\end{bmatrix}.
\end{equation}

Linear stability analysis for $e_1$ is done by solving the corresponding characteristic equation to obtain the following pair of eigenvalues:
\begin{eqnarray}
\label{eqn_SIS_wvd_DFE}
(\lambda_1,\lambda_2)|_{e_1}&{}={}&(0, \beta-\gamma).			%(24)
\end{eqnarray}

The stability of $e_1$ depends on the value taken by $\lambda_2$. The equilibrium point is a stable $DFE$ if $\beta < \gamma$ (or equivalently $\mathcal{R}_0<1$) and unstable if $\beta > \gamma$ (or $\mathcal{R}_0>1$). Similarly, the eigenvalues of the characteristic equation for $e_2$ are:
\begin{eqnarray}
\label{eqn_SIS_wvd_EE}
(\lambda_1,\lambda_2)|_{e_2}&{}={}&(0, -\beta+\gamma).			%(25)
\end{eqnarray}

\begin{figure}[!t]
        \centering
        \begin{subfigure}[b]{0.475\textwidth}
                \centering
                \includegraphics[width=\textwidth]{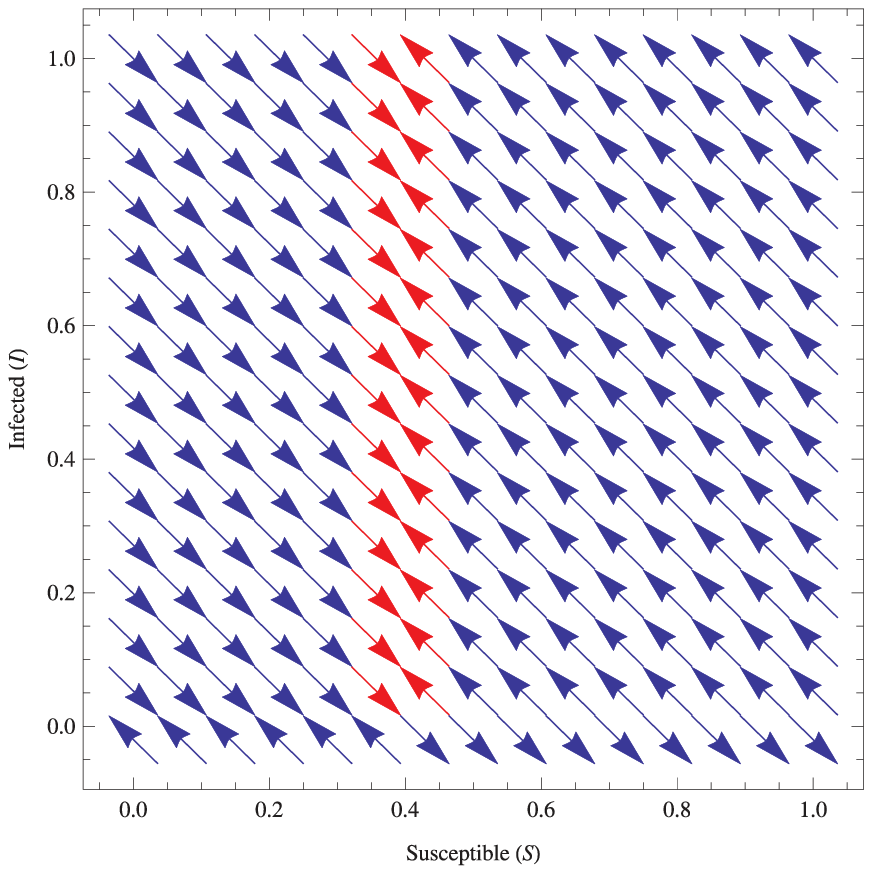}
                \caption{Point $(e_1)$ when $\mathcal{R}_0=2.5$, $\beta=0.5$, and $\gamma=0.2$.}
                \label{Fig10a}
        \end{subfigure}
        ~
        \vspace{0.1in}
        \begin{subfigure}[b]{0.475\textwidth}
                \centering
                \includegraphics[width=\textwidth]{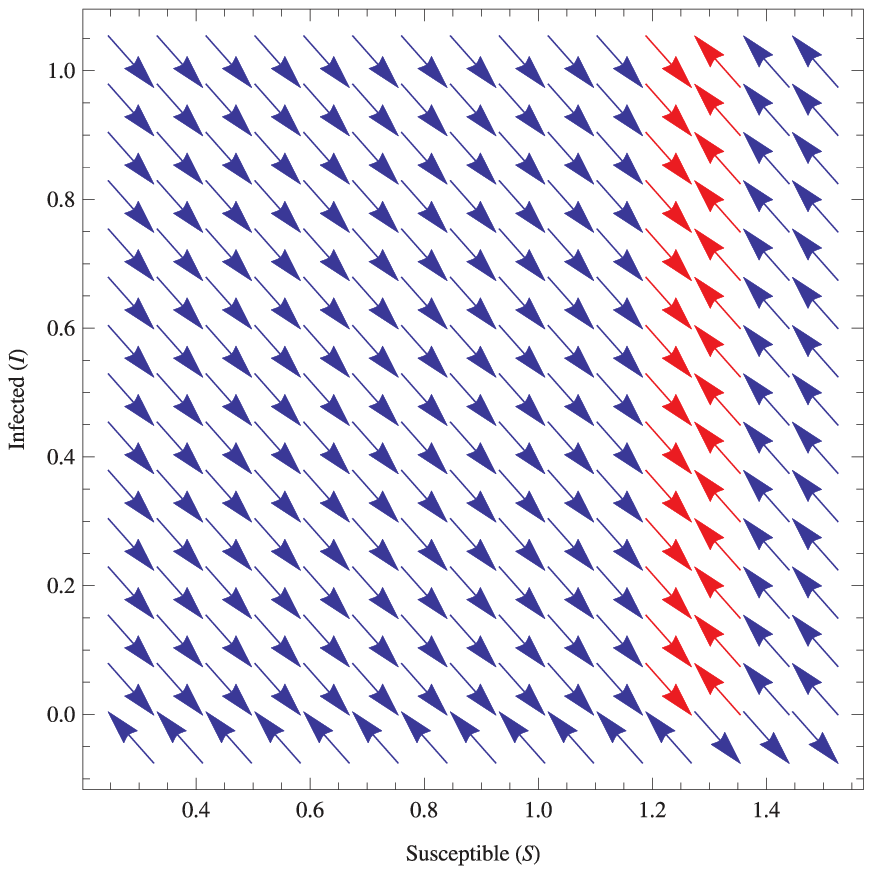}
                \caption{Point $(e_2)$ when $\mathcal{R}_0=0.778$, $\beta=0.7$, and $\gamma=0.9$.}
                \label{Fig10b}
        \end{subfigure}
        \caption[Instability of the equilibrium points in $SIS$  model without vital dynamics]{Vector plots showing the instability of the equilibrium points in $SIS$  model without vital dynamics for (a) $\mathcal{R}_0 > 1$ and ~(b) $\mathcal{R}_0 < 1$.}
        \label{fig10_SIS_wvd_vectorplot}
\end{figure}

In this case, the $EE$ point is stable if $\beta >\gamma$ and unstable if $\beta<\gamma$. Figure~\ref{fig10_SIS_wvd_vectorplot} depicts the vector plots for examples where $e_1$ and $e_2$ are unstable. In Figure~\ref{Fig10a}, the system converges to some state (highlighted in red) other than $(1,0)$ when $\mathcal{R}_0=2.5$. Likewise, for $\mathcal{R}_0<1$, the system converges to an invalid state as illustrated in Figure~\ref{Fig10b}. At $\beta=\gamma$ or equivalently, $\mathcal{R}_0=1$, a \emph{bifurcation} occurs as the two equilibria collide ($DFE$ equals $EE$) and exchange stability \cite{Hethcote2000}. The forward bifurcation occurring at this threshold condition is similar to as seen previously in Section III.

\subsection{SIS Model with vital dynamics}
The $SIS$ model with varying population of constant size is as shown in Figure~\ref{fig11_SIS_vd}. The corresponding system of differential equations for such a model is given below, where $b$ and $\mu$ are assumed to be equal and $S + I = N$:
\begin{eqnarray}
\label{eqn2627_SIS_vd}
\frac{dS}{dt}&{}={}& b N + \gamma I -\beta S \frac{I}{N} - \mu S ,\\		%(26)
\frac{dI}{dt}&{}={}& \beta S \frac{I}{N} - (\gamma + \mu) I.				%(27)
\end{eqnarray}
\begin{figure}[!t]
\centering
\includegraphics[width=1.5in]{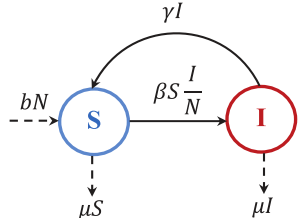}
\caption{The $SIS$ model with vital dynamics.}
\label{fig11_SIS_vd}
\end{figure}

\subsubsection{Existence of equilibria}
\begin{figure*}[!t]
        \centering
        \begin{subfigure}[b]{0.485\textwidth}
                \centering
                \includegraphics[width=\textwidth]{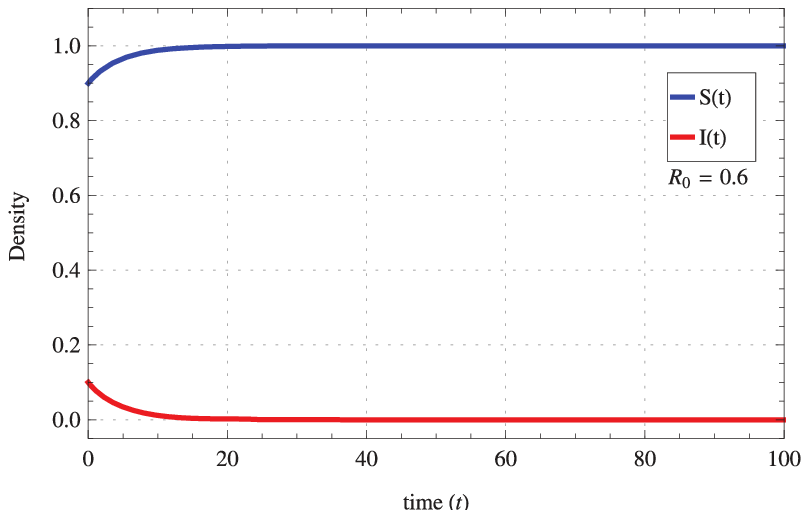}
                \caption{$\mathcal{R}_0=0.6$, $\beta=0.3$, $\gamma=0.3$, and $\mu=0.2$.}
                \label{Fig12a}
        \end{subfigure}
        ~
        \begin{subfigure}[b]{0.48\textwidth}
                \centering
                \includegraphics[width=\textwidth]{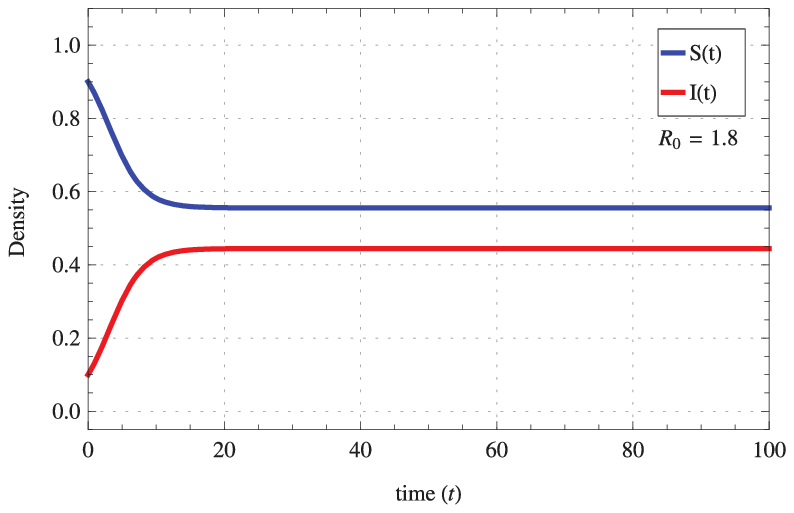}
                \caption{$\mathcal{R}_0=1.8$, $\beta=0.9$, $\gamma=0.3$, and $\mu=0.2$.}
                \label{Fig12b}
        \end{subfigure}%
        \caption[Density versus time for $SIS$ model with vital dynamics]{Density versus time for $SIS$ model with vital dynamics where $N=1$, $S(0)=0.9$ and $I(0)=0.1$.}
        \label{fig12_Densities}
\end{figure*}

To find the equilibrium points of the system, we set (26) and (27) to zero and solve for $S$ and $I$. This results in $e_1$ as the $DFE$ point and $e_2$ as the $EE$ point as given below:
\begin{equation}
\label{eqn28_SIR_vd}
  \begin{split}
   e_1:(S^*,I^*)&=(N, 0),\\		%(28)
e_2:(S^*,I^*)&=\left(\frac{N}{\mathcal{R}_0}, \frac{N}{\mathcal{R}_0}(\mathcal{R}_0-1)\right),
  \end{split}
\end{equation}
where $\mathcal{R}_0$ is $\beta/(\gamma+\mu)$. Figure~\ref{fig12_Densities} shows the system behavior for different values of $\mathcal{R}_0$. In Figure~\ref{Fig12a}, since $\mathcal{R}_0$ is less than 1, the disease dies out and the system enters the disease-free steady state. The same happens at $\mathcal{R}_0 = 1$, where the two equilibria meet. For $\mathcal{R}_0$ greater than 1, the disease does not die out, but instead remains in the population as an endemic with a limiting value. This can be seen in Figure~\ref{Fig12b} where an endemic occurs at $(S(t), I(t))=(0.56, 0.44)$ for $\mathcal{R}_0=1.8$.

\subsubsection{Equilibria stability analysis}
The corresponding Jacobian matrix for this model is given as:
\begin{figure}[!t]
        \centering
        \begin{subfigure}[b]{0.47\textwidth}
                \centering
                \includegraphics[width=\textwidth]{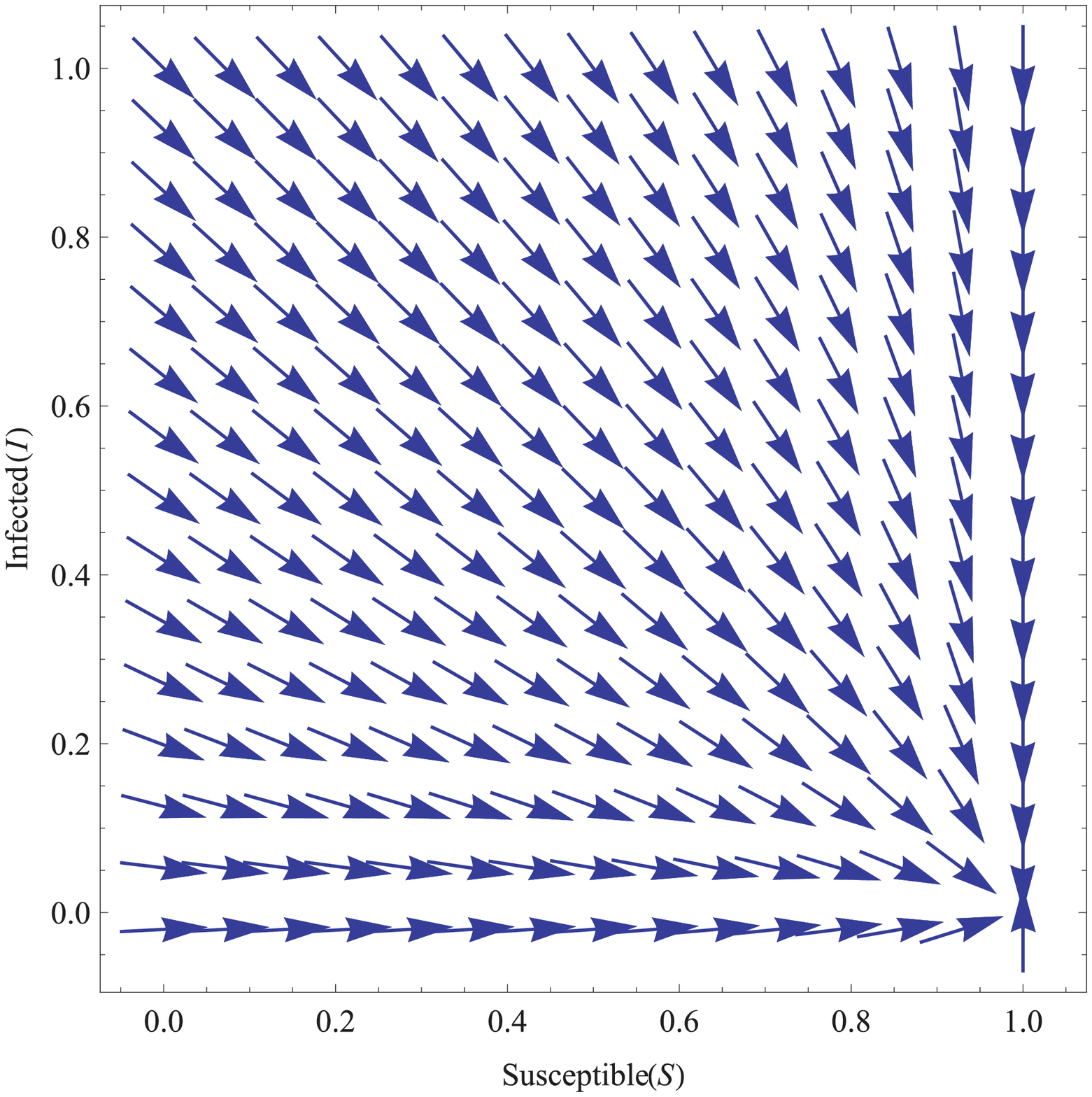}
                \caption{Stability of $e_1$ when $\mathcal{R}_0=0.6$, $\beta=0.3$, $\gamma=0.3$, and $\mu=0.2$.}
                \label{Fig13a}
        \end{subfigure}~
       ~
        \vspace{0.1in}
        \begin{subfigure}[b]{0.47\textwidth}
                \centering
                \includegraphics[width=\textwidth]{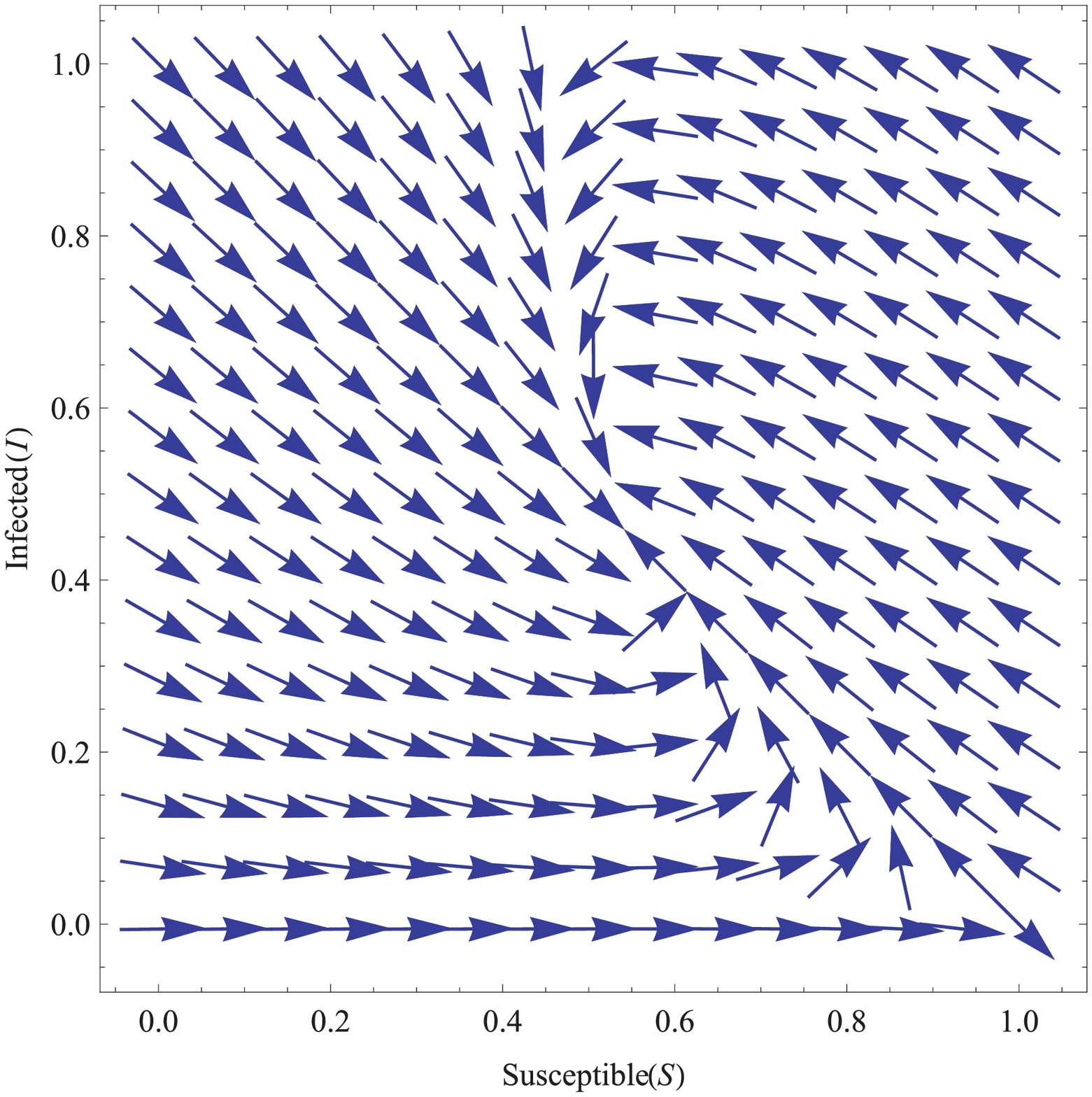}
                \caption{Stability of $e_2$ when $\mathcal{R}_0=1.8$, $\beta=0.9$,  $\gamma=0.9$, and $\mu=0.2$.}
                \label{Fig13b}
        \end{subfigure}
        \caption[Stability of the equilibrium points in $SIS$  model with vital dynamics]{Vector plots showing the stability of the equilibrium points in $SIS$  model with vital dynamics for (a) $\mathcal{R}_0 \leq 1$ and (b) $\mathcal{R}_0 > 1$.}
        \label{fig13_SIS_vd_vectorplot}
\end{figure}
\renewcommand{\arraystretch}{2.0}
\begin{equation}
\label{eqn29_SIS_vd}
J=\begin{bmatrix}
		-\beta \dfrac{I}{N} - \mu &\quad \gamma - \beta \dfrac{S}{N}\\			%(29)
         \beta \dfrac{I}{N} &\quad \beta \dfrac{S}{N} - (\gamma+\mu)
\end{bmatrix}.
\end{equation}

The eigenvalues of $J$ in (\ref{eqn29_SIS_vd}) for $e_1$ and $e_2$ are deduced as follows:
\begin{eqnarray}
(\lambda_1,\lambda_2)|_{e_1}&{}={}&(-\mu,~\beta-(\gamma+\mu)),\\			%(30)
(\lambda_1,\lambda_2)|_{e_2}&{}={}&(-\mu,-\beta+\gamma+\mu).				%(31)
\end{eqnarray}

Linear stability analysis reveals that the disease-free equilibrium ($e_1$) is asymptotically stable if $\beta - (\gamma + \mu) \leq 0$ (or $\mathcal{R}_0 \leq 1$) and unstable otherwise \cite{Vargas2011}. Similarly, the endemic steady state is asymptotically stable if $\mathcal{R}_0>1$. Figure~\ref{fig13_SIS_vd_vectorplot} portraits the stability of $e_1$ and $e_2$ for different values of $\mathcal{R}_0$. In Figure~\ref{Fig13a}, with $\mathcal{R}_0=0.6$, the system converges to $e_1=(1,0)$ where it is stable. In the same manner, as in Figure~\ref{Fig13b}, the system eventually ends up at $e_2=(0.56,0.44)$  for $\mathcal{R}_0=1.8$, which is a valid endemic state. The forward bifurcation for the simple $SIS$ model with demographic turnover and $\mathcal{R}_0$ as the bifurcation parameter occurs at $\mathcal{R}_0=1$ as studied earlier. Nevertheless, such models in presence of additional factors reveal interesting behaviors. Some recent works on $SIS$ model that exhibit bifurcations consider factors such as non-constant contact rate having multiple stable equilibria \cite{Driessche2000}, non-linear birth rate \cite{Liu2009}, treatment \cite{Wang2009}, and time delay \cite{Das2009}.

%~~~~~~~~~~~~~~~~~~~~~~~~~~~~~~~~~~~~~~~~~~~~~~~~~~~~~~~~~~~~~~~~~~~~~~~~~~~~~~~~~~~~~~~~~~~~~~~~~~~~~~~~~
%
%
%
%
%
%~~~~~~~~~~~~~~~~~~~~~~~~~~~~~~~~~~~~~~~~~~~~~ SIRS Model ~~~~~~~~~~~~~~~~~~~~~~~~~~~~~~~~~~~~~~~~~~~~~~~~
\section{The $SIRS$ model}
The $SIRS$ model is an extension of the basic $SIR$ model in which individuals recover with immunity to the disease and become susceptible again after some time recovering. Influenza is a contagious viral disease that is usually studied using this model. In what follows, we investigate the model in both, absence and presence of demographic turnover.

\subsection{$SIRS$ Model without vital dynamics}
\begin{figure}[!t]
\centering
\includegraphics[width=2.0in]{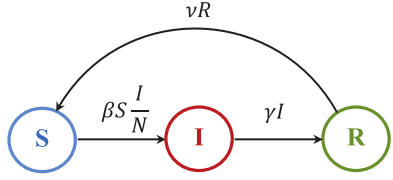}
\caption{The $SIRS$ model without vital dynamics.}
\label{fig14_SIRS_wvd}
\end{figure}
The system of differential equations describing the $SIRS$ flow diagram in Fig.~\ref{fig14_SIRS_wvd} is as below:
\begin{eqnarray}
\label{eqn181920_SIRS_wvd}
\frac{dS}{dt}&{}={}&\nu R - \beta S \frac{I}{N},\\			%(32)
\frac{dI}{dt}&{}={}& \beta S \frac{I}{N}-\gamma I,\\		%(33)
\frac{dR}{dt}&{}={}& \gamma I - \nu R.						%(34)
\end{eqnarray}

\subsubsection{Existence of equilibria}
With the three compartments summing up to $N$, we obtain the following two equilibrium points where $e_1$ and $e_2$ denote the $DFE$ and $EE$, respectively. With  $c_1=\nu/(\gamma+\nu)$, $c_2=\gamma/(\gamma+\nu)$, and $\mathcal{R}_0$ defined as $\beta / \gamma$, we have:
\begin{equation}
\label{eqn35_SIRS_wvd}
  \begin{split}
   e_1:(S^*,I^*,R^*)&=(N, 0, 0),\\			%(35)
e_2:(S^*,I^*,R^*)&=\left(\frac{N}{\mathcal{R}_0}, \frac{N}{\mathcal{R}_0} c_1 (\mathcal{R}_0-1), \frac{N}{\mathcal{R}_0} c_2(\mathcal{R}_0-1)\right).
  \end{split}
\end{equation}

\begin{figure}[t]
        \centering
        \begin{subfigure}[b]{0.485\textwidth}
                \centering
                \includegraphics[width=\textwidth]{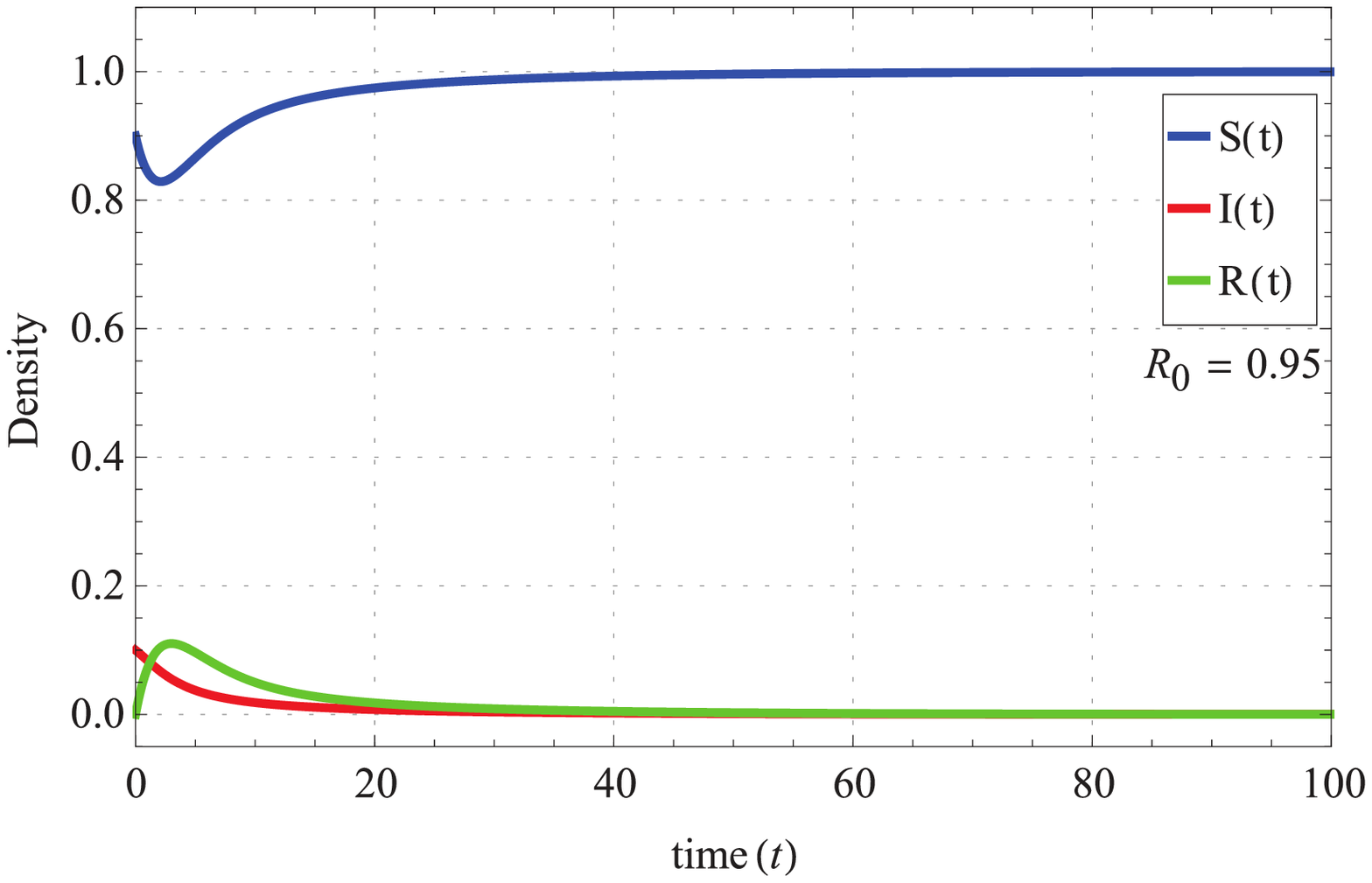}
                \caption{$\mathcal{R}_0=0.95$, $\beta=0.95$, $\gamma=1$, and $\nu=0.5$.}
                \label{Fig15a}
        \end{subfigure}~
        ~
        \begin{subfigure}[b]{0.485\textwidth}
                \centering
                \includegraphics[width=\textwidth]{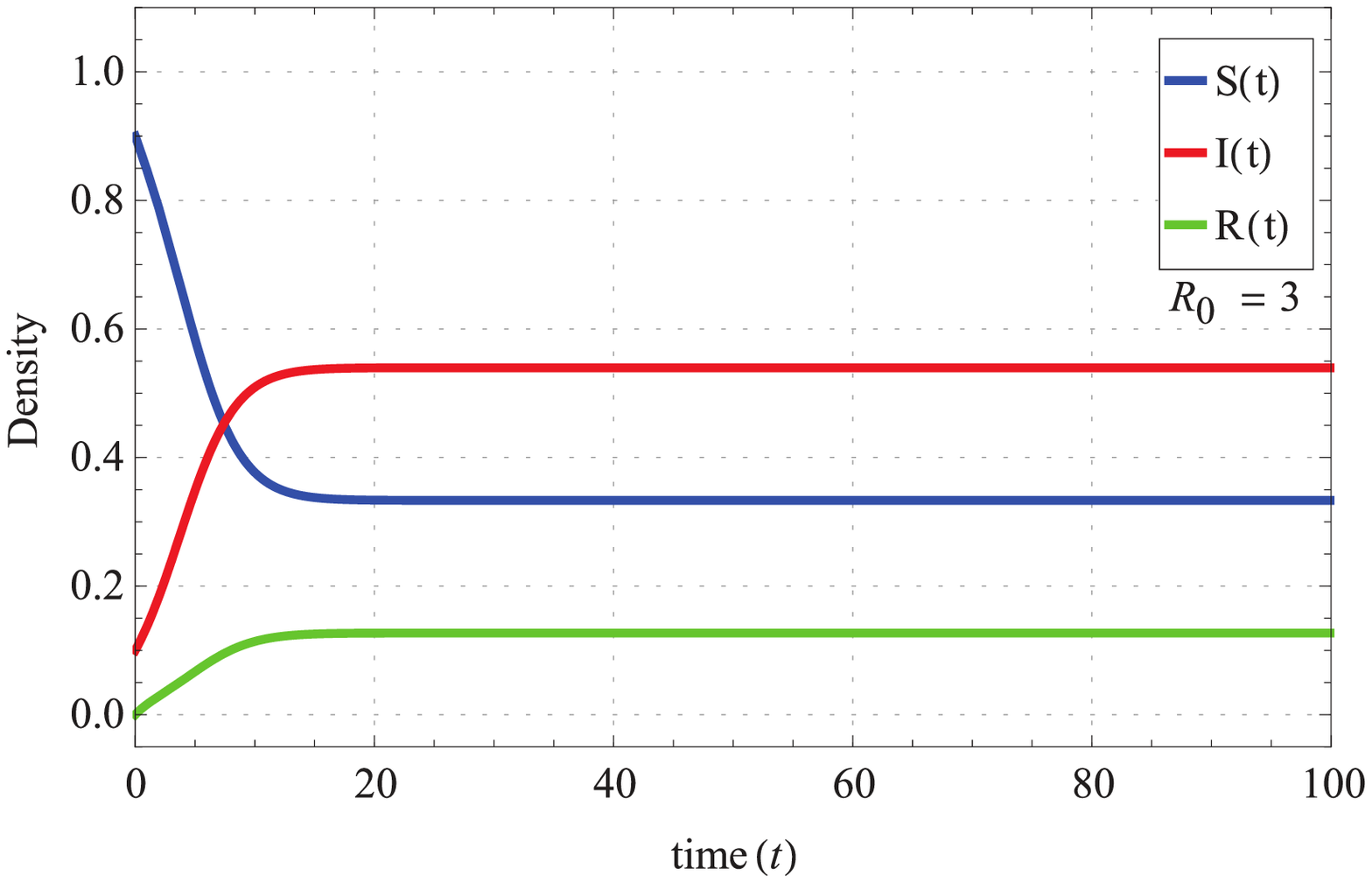}
                \caption{$\mathcal{R}_0=3$, $\beta=0.6$, $\gamma=0.2$, and $\nu=0.85$.}
                \label{Fig15b}
        \end{subfigure}
        \caption[Density versus time for $SIRS$ model without vital dynamics]{Density versus time for $SIRS$ model without vital dynamics where $N=1$, $S(0)=0.9$, $I(0)=0.1$, and $R(0)=0$.}
        \label{fig15_Densities}
\end{figure}

As illustrated in Fig.~\ref{fig15_Densities}, the infection dies out and reaches the disease-free steady state for $\mathcal{R}_0 \leq 1$ and stays as an endemic when $\mathcal{R}_0>1$.

\subsubsection{Equilibria stability analysis}
The corresponding Jacobian matrix of the system is obtained by substituting $R$ with $N-S-I$ in (32). Hence,
\begin{equation}
\label{eqn36_SIRS_wvd}
J=\begin{bmatrix}
		-\nu - \beta \dfrac{I}{N} &~~ -\nu - \beta \dfrac{S}{N}\\			%(36)
         \beta \dfrac{I}{N} &~~ \beta \dfrac{S}{N} - \gamma
\end{bmatrix}.
\end{equation}

At $e_1$, the matrix $J$ yields the following two eigenvalues:
\begin{eqnarray}
(\lambda_1,\lambda_2)|_{e_1}=(-\nu,~\beta-\gamma).				%(37)
\end{eqnarray}

Therefore, $e_1$ is a stable node if $\lambda_2 \leq 0$ or $\mathcal{R}_0 \leq 1$, and is a saddle point (unstable) if $\lambda_2 > 0$ or $\mathcal{R}_0>1$. However, analyzing the stability of $e_2$ requires more care due to the structure complexity of the corresponding eigenvalues given as:
\begin{equation}
\label{eqn38_SIRS_wvd}
  \begin{split}
	\lambda_{1,2}|_{e_2}&=\frac{-\nu (\beta+\nu)\pm \sqrt{\nu^2 (\beta + \nu)^2-4 \nu(\beta-\gamma)(\gamma+\nu)^2}}{2(\gamma+\nu)}.			%(38)
  \end{split}
\end{equation}

For the sake of clarity, let us denote $\nu(\beta+\nu)$ and $2(\gamma+\nu)$ as $c$ and $b$, respectively. Rewriting the above eigenvalues gives:
\begin{eqnarray}
\lambda_{1,2}|_{e_2}=\frac{-c\pm \sqrt{c^2 - \nu(\beta-\gamma)b^2}}{b}.			%(39)
\end{eqnarray}

Also, let us represent $c^2-\nu (\beta-\gamma) b^2$ as $\delta$ and $\nu (\beta-\gamma) b^2$ as $\zeta$. In order to study the stability of $e_2$, we need to consider the following two cases:
\begin{figure*}[!t]
        \centering
        \begin{subfigure}[b]{0.47\textwidth}
                \centering
                \includegraphics[width=\textwidth]{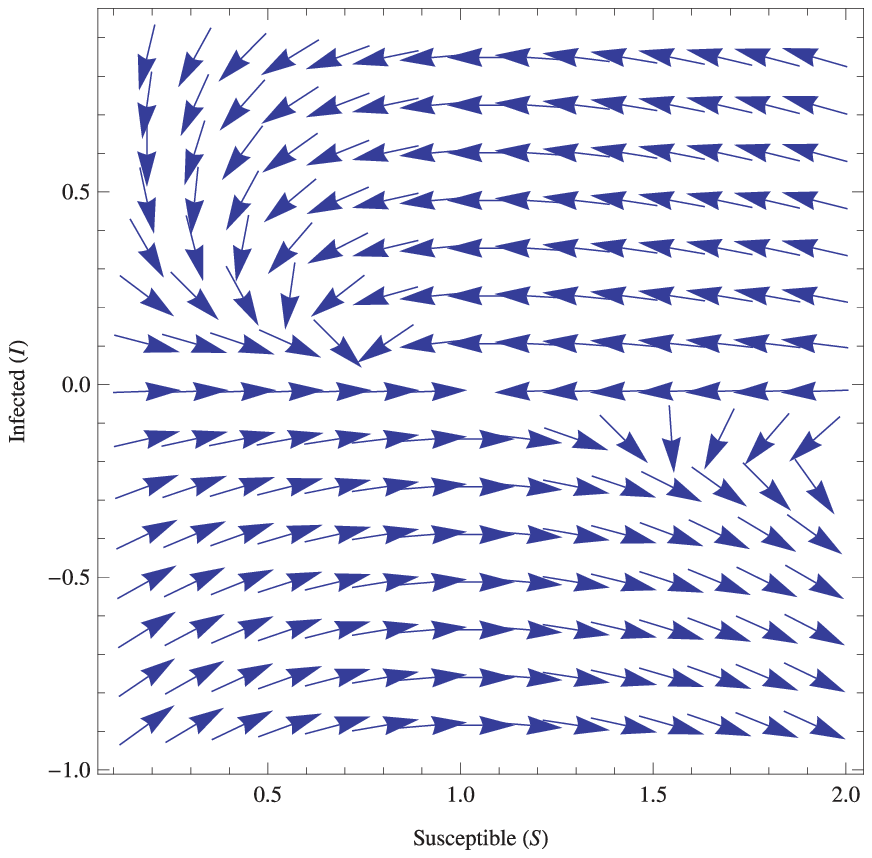}
                \caption{Saddle point with $\mathcal{R}_0=0.95$, $\beta=0.95$, $\gamma=1$, and $\mu=0.5$.}
                \label{Fig16a}
        \end{subfigure}~
        ~
        \vspace{0.1in}
        \begin{subfigure}[b]{0.47\textwidth}
                \centering
                \includegraphics[width=\textwidth]{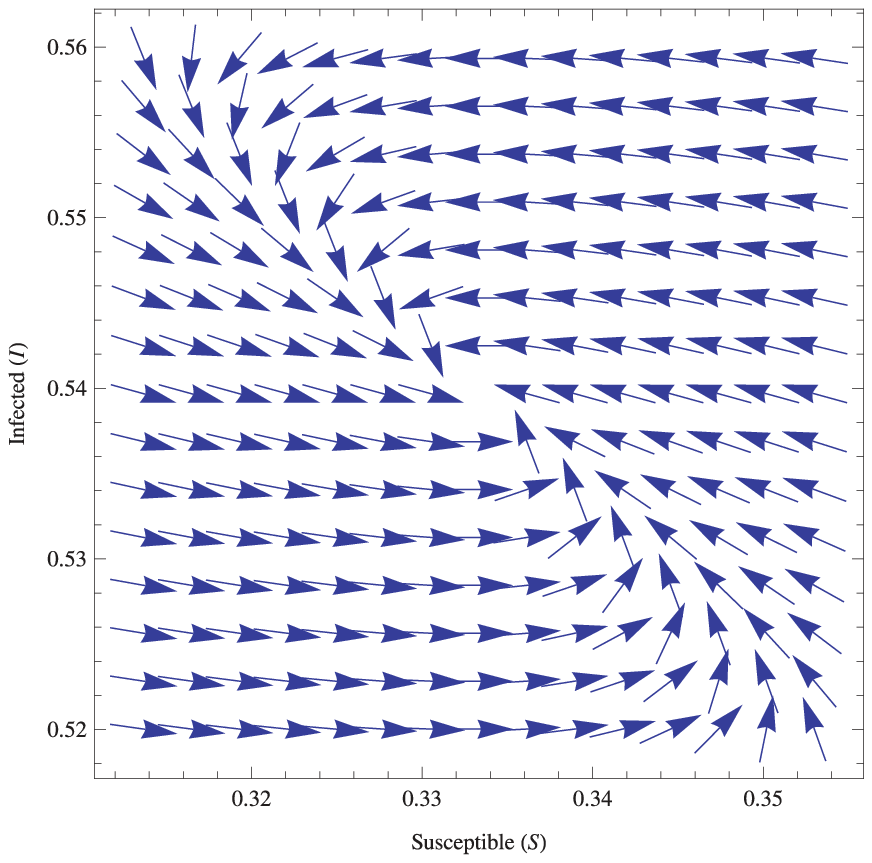}
                \caption{Stable node with $\mathcal{R}_0=3$, $\beta=0.6$, $\gamma=0.2$, and $\mu=0.85$.}
                \label{Fig16b}
        \end{subfigure}
        
        ~
        \begin{subfigure}[b]{0.47\textwidth}
                \centering
                \includegraphics[width=\textwidth]{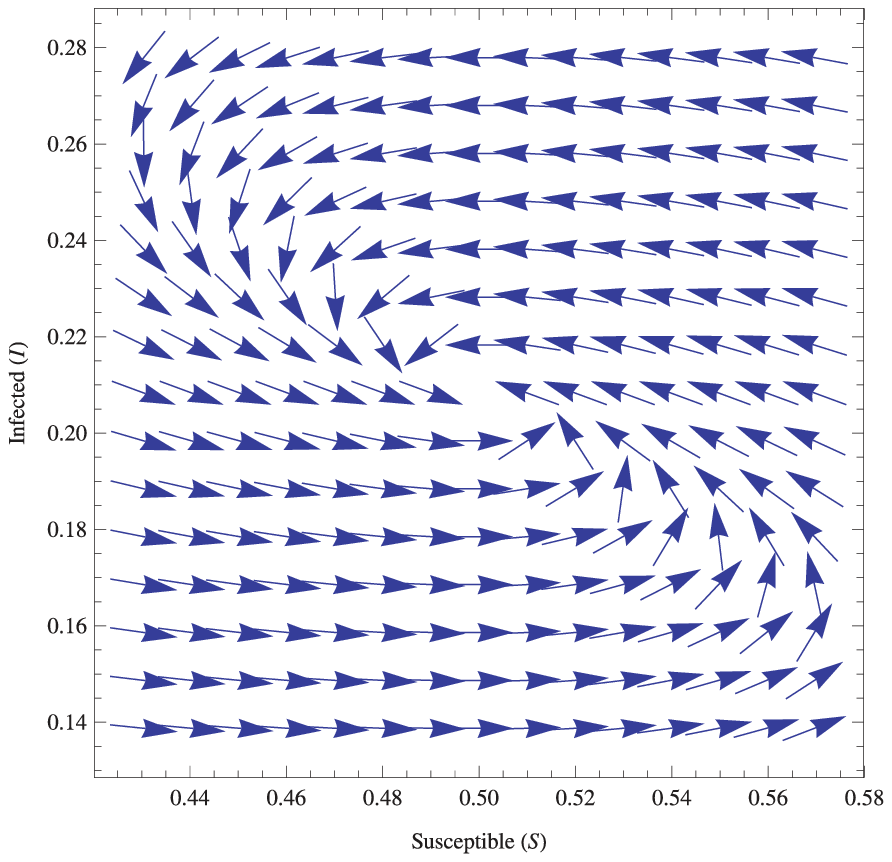}
                \caption{Stable focus with $\mathcal{R}_0=2$, $\beta=0.7$, $\gamma=0.35$, and $\mu=0.25$.}
                \label{Fig16c}
        \end{subfigure}
        \caption[Stability of the $EE$ point in $SIRS$  model without vital dynamics]{Vector plots showing the stability of the $EE$ point in $SIRS$  model without vital dynamics for (a) $\delta \geq 0$ and $\mathcal{R}_0 \leq 1$ ($Case~A$ (i)), (b) $\delta \geq 0$ and $\mathcal{R}_0 > 1$ ($Case~A$ (ii)), and (c) $\delta < 0$ and $\mathcal{R}_0 > 1$ ($Case~B$).}
        \label{fig16_SIRS_portrait}
\end{figure*}
\begin{enumerate}[~$\bullet$]
	\item \emph{Case A:} When $\delta \geq 0$, depending on the possible values that $(\beta-\gamma)$ can take, we have the following two conditions :
	\begin{enumerate}[(i)]
		\item If $\beta - \gamma \leq 0$ or $\mathcal{R}_0 \leq 1$, then $\zeta \leq 0$ which implies that the magnitude of $\delta$ is always greater than or equal to $c^2$. Under this condition, the eigenvalues will always be real with opposite signs. Hence, the equilibrium point $e_2$ would be a \emph{saddle point} as portrayed in Figure~\ref{Fig16a}, where the parametric values are the same as that in Figure~\ref{Fig15a}.
		\item If $\beta - \gamma > 0$ or $\mathcal{R}_0 > 1$, then $\zeta > 0$ and thus, the magnitude of $\delta$ would always be lesser than $c^2$. In this case, $\lambda_1$ and $\lambda_2$ will always be real values with negative signs. As shown in Figure~\ref{Fig16b}, $e_2$ converges to a \emph{stable node} with the same parametric values as given in Figure~\ref{Fig15b}. Here, $\delta=0.019$ which is lesser than $c^2=1.519$ and thus, results in a stable node at $(0.333, 0.539)$.
	\end{enumerate}
	\item \emph{Case B:} When $\delta < 0$, $\beta - \gamma$ should always be greater than zero and thus, the eigenvalues would be complex conjugates. Since the real parts of the eigenvalues are negative, the equilibrium would be a \emph{stable focus} point where the following condition holds:
\setlength{\arraycolsep}{0.0em}
\begin{eqnarray}
	\mathcal{R}_0 > \frac{c^2}{\gamma \nu b^2} + 1.			%(40)
	\label{eqn40_SIRS_wvd}
\end{eqnarray}

As an example for this case, with $\mathcal{R}_0=2$, $\beta=0.7$, $\gamma=0.35$, and $\nu=0.25$, the stable focus occurs at $(0.5, 0.208)$ as depicted in Figure~\ref{Fig16c}.
\end{enumerate}

\subsection{$SIRS$ model with vital dynamics}
\begin{figure}[!t]
\centering
\includegraphics[width=2.3in]{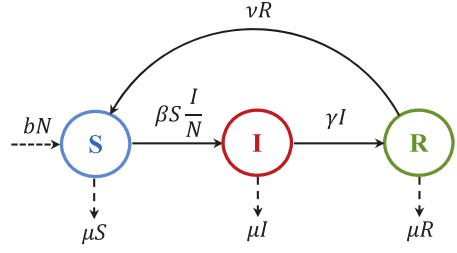}
\caption{The $SIRS$ model with vital dynamics.}
\label{fig17_SIRS_vd}
\end{figure}
As in Figure~\ref{fig17_SIRS_vd}, the $SIRS$ model with standard incidence can be simply expressed as the following set of differential equations \cite{Vargas2011}:
\begin{eqnarray}
\label{eqn414243_SIRS_vd}
\frac{dS}{dt}&{}={}&b N + \nu R - \beta S \frac{I}{N} - \mu S,\\			%(41)
\frac{dI}{dt}&{}={}& \beta S \frac{I}{N} - (\gamma + \mu) I,\\				%(42)
\frac{dR}{dt}&{}={}& \gamma I - (\nu + \mu) R.								%(43)
\end{eqnarray}

It is worth mentioning that $1/\gamma$ and $1/\nu$ can be regarded as the \emph{mean infectious period} and the \emph{mean immune period}, respectively. With $\nu =0$, the model reduces to an $SIR$ model with no transition from class $R$ to class $S$ due to life-long immunity.

\subsubsection{Existence of equilibria}
\begin{figure}[!t]
        \centering
        \begin{subfigure}[b]{0.485\textwidth}
                \centering
                \includegraphics[width=\textwidth]{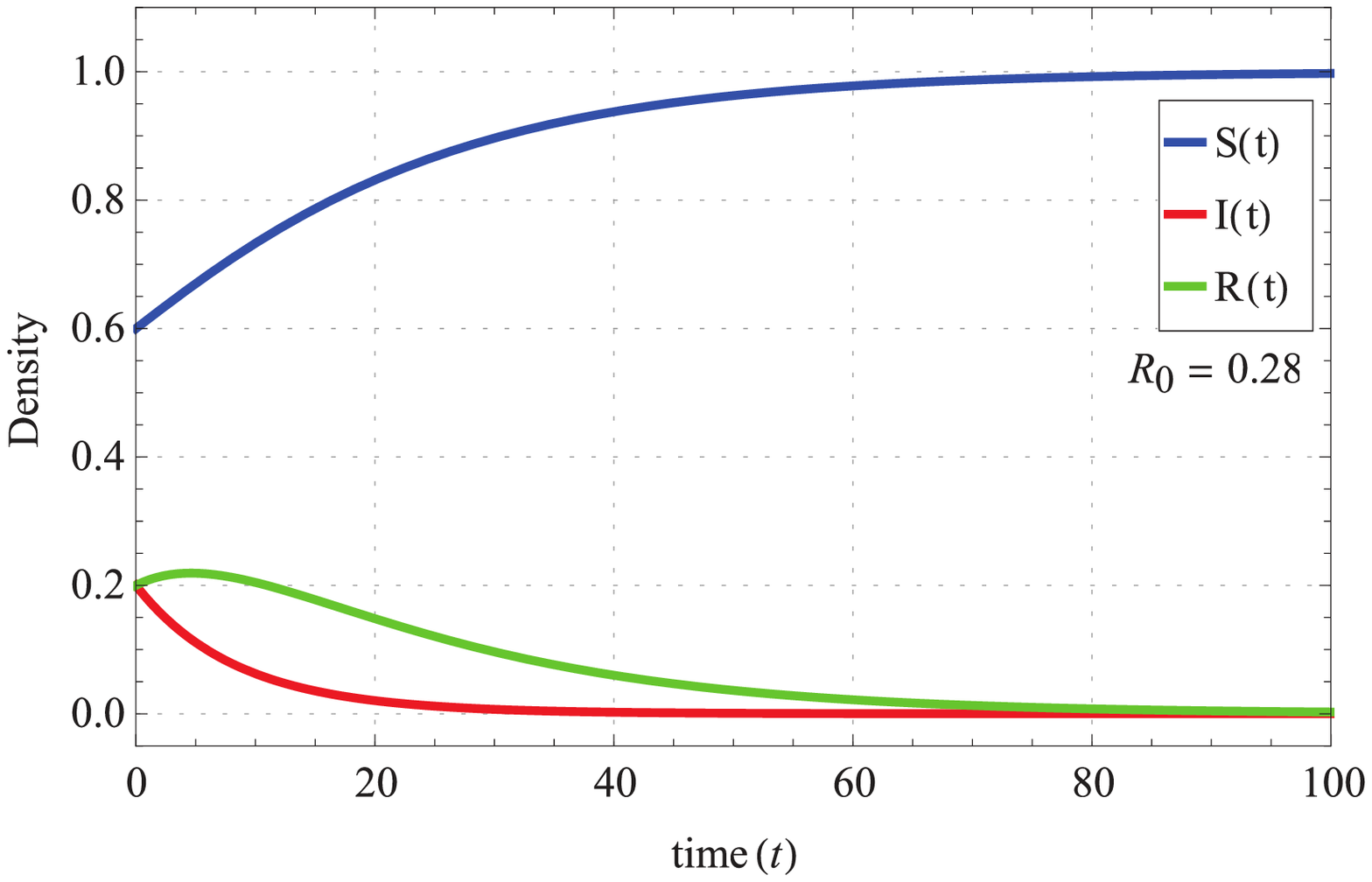}
                \caption{$\mathcal{R}_0=0.28$, $\beta=0.04$, $\gamma=0.1$, $\mu=0.043$, and $\nu=0.01$.}
                \label{Fig18a}
        \end{subfigure}    
        ~
        \begin{subfigure}[b]{0.485\textwidth}
                \centering
                \includegraphics[width=\textwidth]{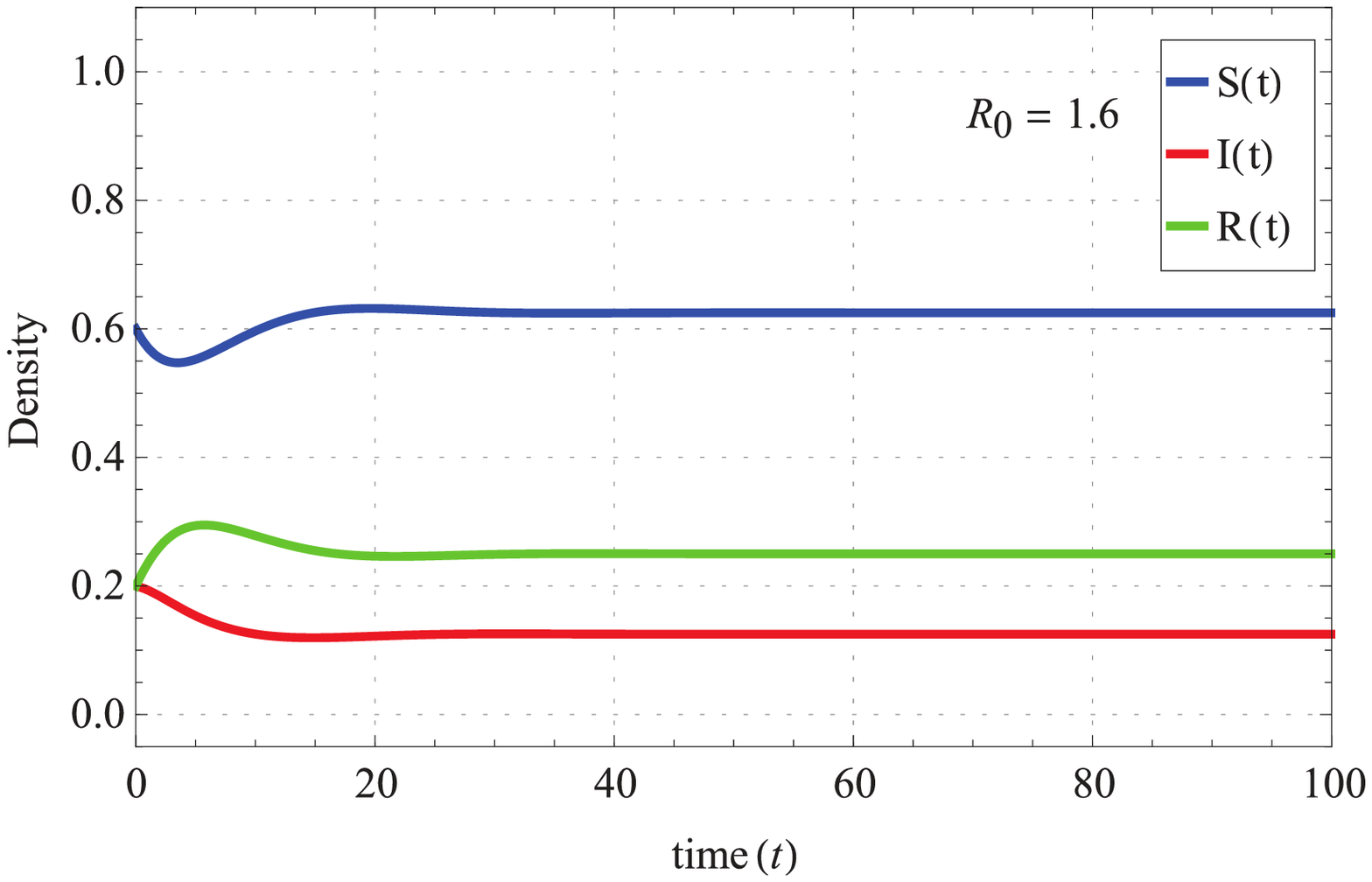}
                \caption{$\mathcal{R}_0=1.6$, $\beta=0.8$, $\gamma=0.4$, $\mu=0.1$, and $\nu=0.1$.}
                \label{Fig18b}
        \end{subfigure}%
        \caption[Density versus time for $SIRS$ model with vital dynamics]{Density versus time for $SIRS$ model with vital dynamics where $N=1$, $S(0)=0.6$, $I(0)=0.2$, and $R(0)=0.2$.}
        \label{fig18_Densities}
\end{figure}

The system has a $DFE$ point and a unique $EE$ point denoted by $e_1$ and $e_2$, respectively, as given below:
\begin{equation}
\label{eqn44_SIRS_vd}
  \begin{split}
   e_1:(S^*,I^*,R^*)&=(N, 0, 0),\\
e_2:(S^*,I^*,R^*)&=\left(\frac{N}{\mathcal{R}_0}, \frac{N}{\mathcal{R}_0} c_1 (\mathcal{R}_0-1), \frac{N}{\mathcal{R}_0} c_2(\mathcal{R}_0-1)\right),			%(44)
  \end{split}
\end{equation}
where $\mathcal{R}_0$ is given by $\beta/(\gamma+\mu)$, $c_1=(\nu+\mu)/(\gamma+\nu+\mu)$, $c_2=\gamma/(\gamma+\nu+\mu)$, and $b$ is assumed to be equal to $\mu$. It should be noted that the basic reproduction number does not depend on the loss of immunity rate ($\nu$). The system reaches $DFE$ steady state for $\mathcal{R}_0 =0.28$ as shown in Figure~\ref{Fig18a}. Here, the parametric values are taken to be $\beta=0.04$, $\mu=0.043$, $\gamma=0.1$, and $\nu=0.01$. On the other hand, Figure~\ref{Fig18b} illustrates an example for which $\mathcal{R}_0 > 1$. In this case, the system reaches the endemic state $(0.625,0.125,0.25)$ for $\mathcal{R}_0 = 1.6$.

\subsubsection{Equilibria stability analysis}
Substituting $R=N-S-I$ in (41) gives the following Jacobian matrix which is obtained from (41) and (42):
\begin{equation}
\label{eqn45_SIRS_vd}
J=\begin{bmatrix}
		- \nu - \beta \dfrac{I}{N} - \mu &~~ - \nu - \beta \dfrac{S}{N}\\
         \beta \dfrac{I}{N} &~~ \beta \dfrac{S}{N}-(\gamma+\mu)					%(45)
\end{bmatrix}.
\end{equation}

Evaluating (45) at $e_1$ and solving its corresponding characteristic equations yields the following pair of eigenvalues:
\begin{eqnarray}
(\lambda_1,\lambda_2)|_{e_1}&{}={}&(-(\nu+\mu),~\beta-(\gamma+\mu)).			%(46)
\end{eqnarray}

\begin{figure}[!t]
        \centering
        \begin{subfigure}[b]{0.46\textwidth}
                \centering
                \includegraphics[width=\textwidth]{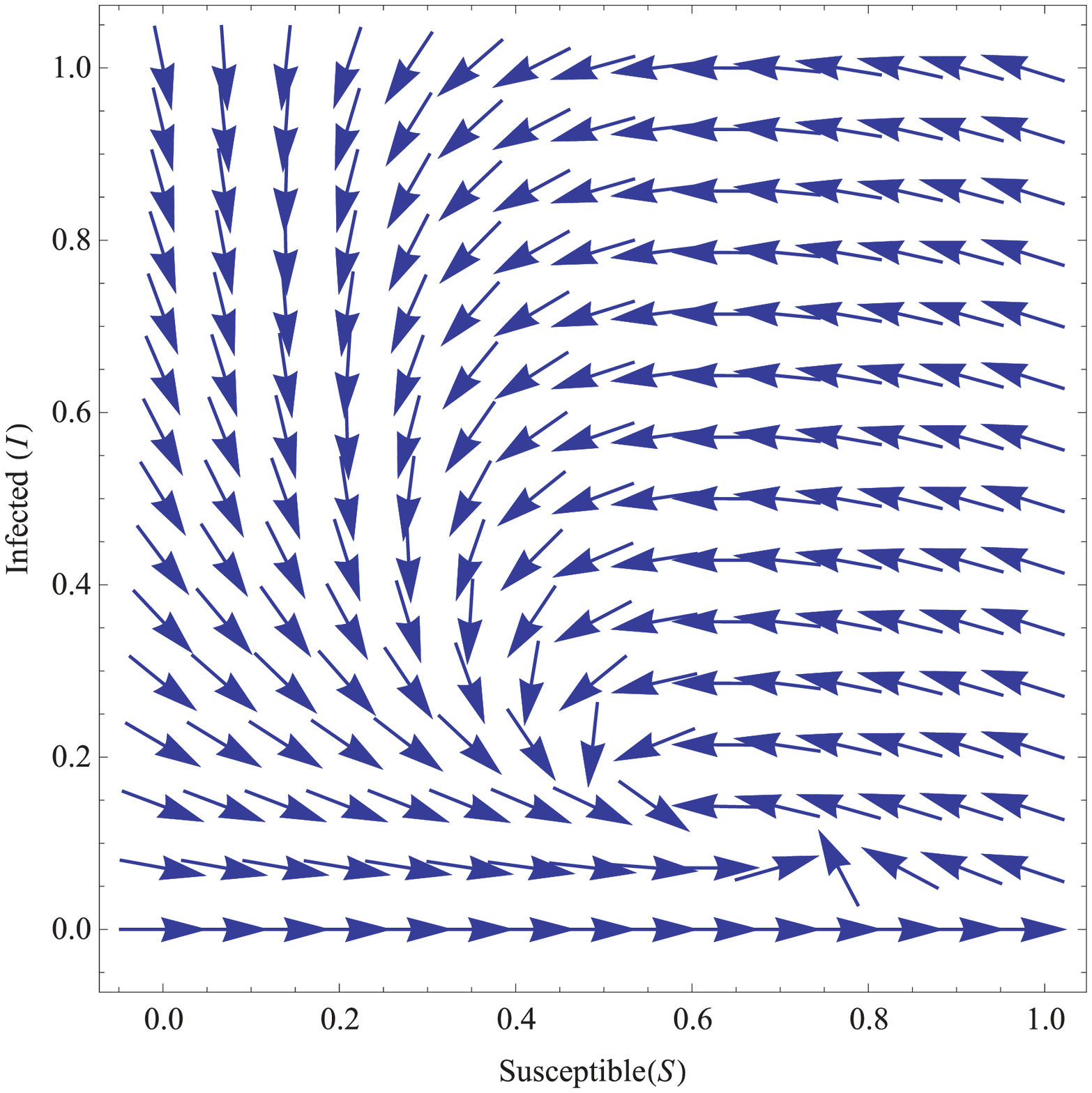}
                \caption{Instability of $DFE$ point when $\mathcal{R}_0=1.6$,  $\beta = 0.8$, $\gamma = 0.4$, $\mu = 0.1$, and $\nu = 0.1$.}
                \label{Fig19a}
        \end{subfigure}~
        ~
        \vspace{0.1in}
        \begin{subfigure}[b]{0.47\textwidth}
                \centering
                \includegraphics[width=\textwidth]{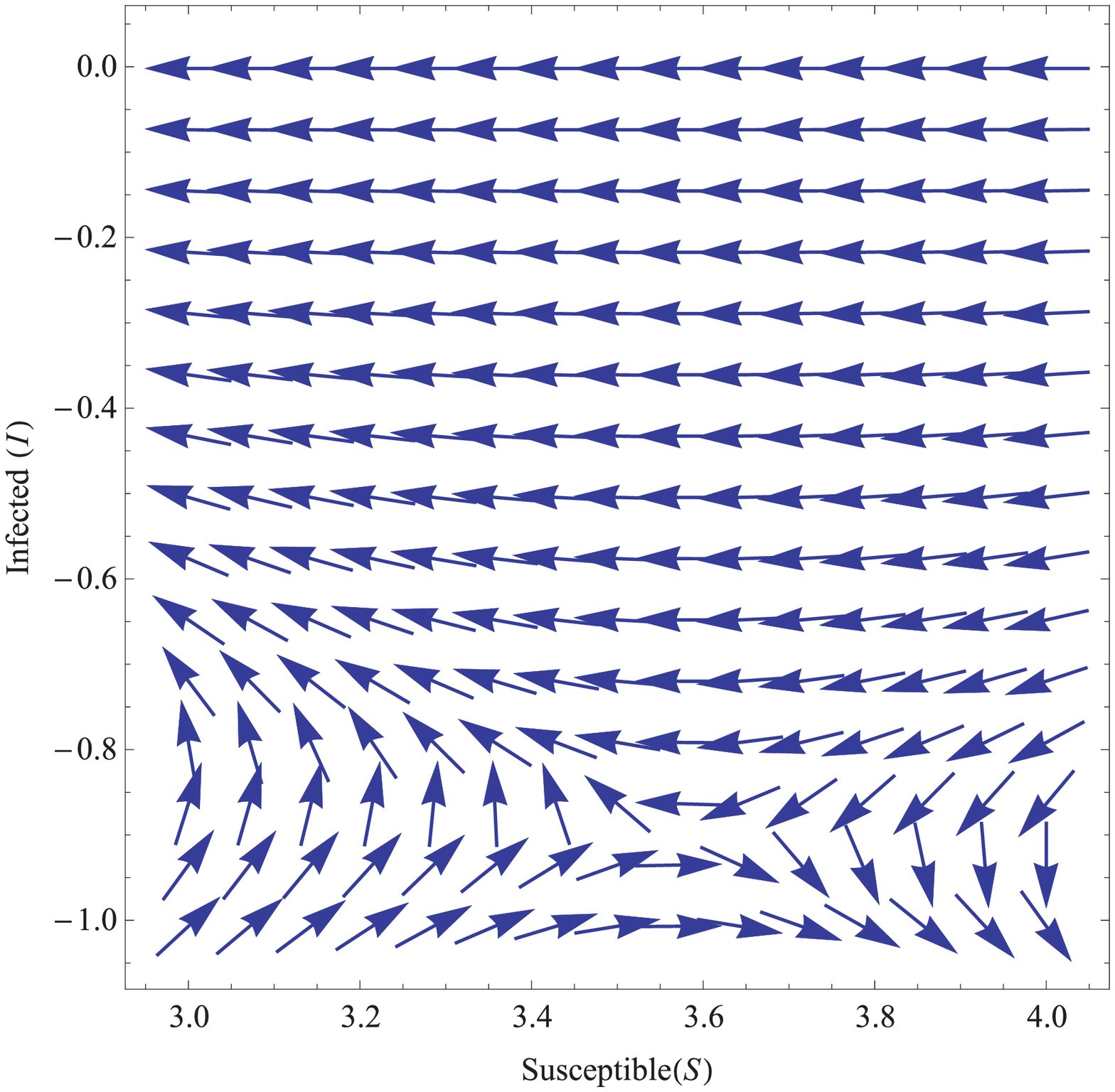}
                \caption{Instability of $EE$ point when $\mathcal{R}_0=0.28$, $\beta = 0.04$, $\gamma=0.1$, $\mu = 0.043$, and $\nu = 0.01$.}
                \label{Fig19b}
        \end{subfigure}
        \caption[Instability of the equilibrium points in $SIRS$ model with vital dynamics]{Vector plots showing the instability of $SIRS$ model with vital dynamics for (a) disease-free equilibrium when $\mathcal{R}_0 > 1$  and ~(b) endemic equilibrium  when  $\mathcal{R}_0 \leq 1$.}
        \label{fig19_SIRS_portrait}
\end{figure}

We see that the disease-free equilibrium ($e_1$) is stable if $\lambda_2 \leq 0$, i.e. $\beta-(\gamma+\mu) \leq 0$ or $\mathcal{R}_0 \leq 1$, and unstable otherwise. Similarly, on finding the eigenvalues for $e_2$, we see that this unique endemic equilibrium point is stable for $\mathcal{R}_0 > 1$. The instability of the equilibrium points can be seen in Figure~\ref{fig19_SIRS_portrait}. For $\mathcal{R}_0 = 1.6$, the vector plot in Figure~\ref{Fig19a} shows how the system does not converge to $(S^*,I^*)=(1,0)$. In the same manner, Figure~\ref{Fig19b} illustrates the instability of $e_2$ at $(S^*,I^*)=(3.575,-0.892)$ when $\mathcal{R}_0$ is less than unity. With the basic reproduction ratio as the bifurcation parameter, the system yields a transcritical forward bifurcation at $\mathcal{R}_0=1$. More interesting behaviors have been reported when studied under factors such as stage structure \cite{Zhang2010} and non-linear incidence rates \cite{Alexander2006}, \cite{Hu2011}.

Hitherto, we have dealt with models comprising of $S$, $I$, and $R$ compartments. In these models, the infected individuals become infectious immediately. In the next two models, an \emph{exposed} compartment in which all the individuals have been infected but are not yet infectious, is introduced. Such models take into consideration the latent period of the disease, resulting in an additional compartment denoted by $E(t)$. The progression rate coefficient from compartment $E$ to $I$ is given as $\varepsilon$ such that $1/\varepsilon$ is the mean latent period. Several other models with latent period such as $SEIR$ and $SEIRS$ have also been reported in the literature. However, such models are beyond the scope of this article. Interested readers can refer to \cite{Keeling2011}, \cite{Vynnycky2010}, and \cite{Ma2009} for more on epidemic models beyond two dimensions.

%~~~~~~~~~~~~~~~~~~~~~~~~~~~~~~~~~~~~~~~~~~~~~~~~~~~~~~~~~~~~~~~~~~~~~~~~~~~~~~~~~~~~~~~~~~~~~~~~~~~~~~~~~
%
%
%
%
%
%~~~~~~~~~~~~~~~~~~~~~~~~~~~~~~~~~~~~~~~~~~~~~~ SEI Model ~~~~~~~~~~~~~~~~~~~~~~~~~~~~~~~~~~~~~~~~~~~~~~~~
\section{The $SEI$ model}
Unlike $SIR$ models, the \emph{susceptible-exposed-infected} ($SEI$) model assumes that a susceptible individual first undergoes a latent (or exposed) period before becoming infectious \cite{Anderson1981}. One example of this model is the transmission of severe acute respiratory syndrome (SARS) coronavirus.
\begin{figure}[!t]
	\centering
	\includegraphics[width=2.0in]{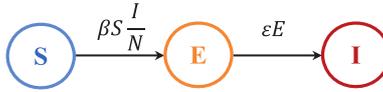}
	\caption{The $SEI$ model without vital dynamics.}
	\label{fig20_SEI_wvd}
\end{figure}

\subsection{$SEI$ model without vital dynamics}
For a fixed population of size $N$, the following differential equations describe the flow diagram in Fig.~\ref{fig20_SEI_wvd}:
\begin{eqnarray}
\label{eqn484950_SEI_wvd}
\frac{dS}{dt}&{}={}& - \beta S \frac{I}{N},\\					%(47)
\frac{dE}{dt}&{}={}& \beta S \frac{I}{N} - \varepsilon E,\\		%(48)
\frac{dI}{dt}&{}={}& \varepsilon E.								%(49)
\end{eqnarray}

The system should be analyzed asymptotically as it does not have a closed-form solution. We shall see that the population converges into a single compartment due to the straight-forward nature of the system. In what follows, we investigate the system behavior in terms of $\beta$ and $\varepsilon$.
\begin{enumerate}[~$\bullet$]
	\item \emph{Case A:} When $\beta\neq 0$ and $\varepsilon\neq 0$, depending on the initial values of $E(0)$ and $I(0)$, the system approaches two different equilibrium points as below:
		\begin{enumerate}[(i)]
			\item If $E(0)=0$ and $I(0)=0$, the system remains in the following disease-free equilibrium:
					\begin{eqnarray}
						(S(t),E(t),I(t))=(N,0,0).			%(50)
					\end{eqnarray}	
				This scenario is illustrated in Figure~\ref{Fig21a} where the complete population is susceptible at all times for $N=1$, $\beta=0.8$, $\varepsilon=0.5$, and $S(0)=1$.
			\item If $E(0)\neq0$ or $I(0)\neq0$, then the system approaches the following equilibrium point in long-term:
					\begin{eqnarray}
						\lim_{t \to \infty}(S(t),E(t),I(t))=(0,0,N).		%(51)
					\end{eqnarray}
				To prove this, consider the solution of (49) which is:
					\begin{eqnarray}
						I(t)=\varepsilon \int_0^t E(\tau) d\tau+I(0).		%(52)
					\end{eqnarray}
				$I(t)$ is a monotonically increasing function when $E(t)>0$. Since all compartments are always non-negative, $E(t)$ is greater than 0 when $E(t)\neq0$. Additionally, on solving (47), we get:
					\begin{eqnarray}
						S(t)=S(0) \exp\left[-\frac{\beta}{N}\int_0^t I(\tau) d\tau\right].		%(53)
					\end{eqnarray}
				When $I(t)>0$, we see that $S(t)$ is a monotonically decreasing function. Unless when $E(t)=0$ for all $t$, $I(t)$ is a monotonically increasing function, and $\lim_{t \to \infty} S(t)=0$ as $\lim_{t \to \infty} \int_0^t I(\tau) d\tau=\infty$. In terms of $E(t)$, from (48), we see that the solution $E(t)=c e^{-\varepsilon t}$ goes to 0 as $t$ approaches infinity. In summary, if the condition $E(0)\neq0$ or $I(0)\neq0$ is satisfied, then $\lim_{t \to \infty}S(t)=0$. Since $S=0$, $\lim_{t \to \infty}E(t)=\lim_{t \to \infty}c e^{-\varepsilon t}=0$. Consequently, $\lim_{t \to \infty}I(t)=N$. This case is depicted in Figure~\ref{Fig21b}.
		\end{enumerate}
\item \emph{Case B:} When $\beta=0$ and $\varepsilon\neq0$, the reduced system yields the following solution:	
		\begin{equation}
		\label{eqn55_SEI_wvd}
  			\begin{split}
  			 S(t)&= S(0),\\
			E(t)&= E(0)e^{-\varepsilon t},\\
			I(t)&= \varepsilon \int_0^t E(\tau) d\tau + I(0).			%(54)
  			\end{split}
		\end{equation}		
		Since $\lim_{t \to \infty}E(t)=0$, the system results in the following equilibrium solution as shown in Figure~\ref{Fig21c}, where the state of the system changes from the initial condition $(S(0),E(0),I(0))=(0.75,0.24,0.01)$ to steady state $(0.75,0,0.25)$ for $\varepsilon=0.5$:
		\begin{eqnarray}
			\lim_{t \to \infty}(S(t),E(t),I(t))=(S(0),0,N-S(0)).		%(55)
		\end{eqnarray}
\begin{figure*}[!t]
        \centering
        \begin{subfigure}[b]{0.485\textwidth}
                \centering
                \includegraphics[width=\textwidth]{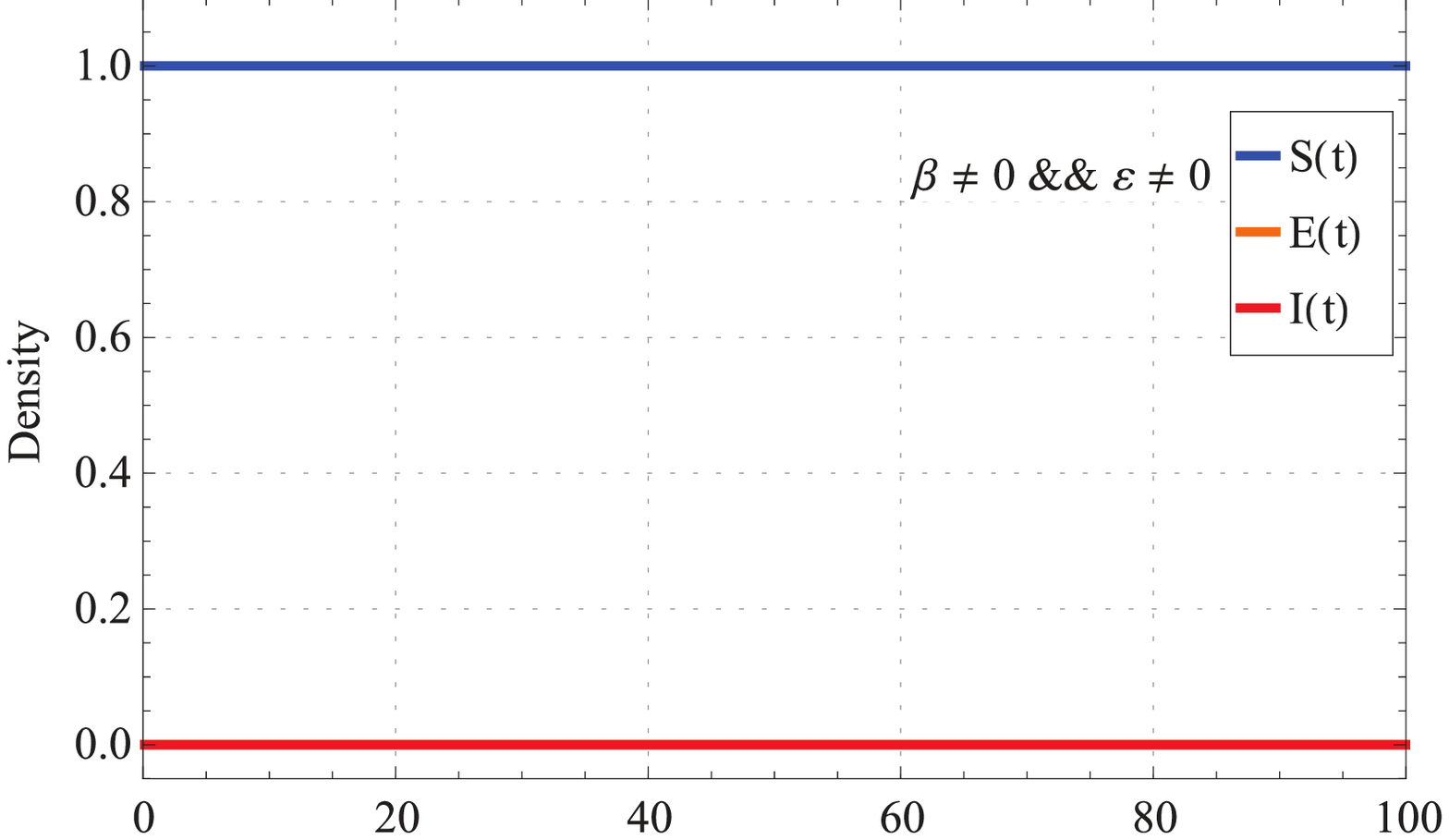}
                \caption{$S(0)=1$, $E(0)=I(0)=0$, $\beta=0.8$, and $\varepsilon=0.5$.}
                \label{Fig21a}
        \end{subfigure}% 
        ~
        \begin{subfigure}[b]{0.485\textwidth}
                \centering
                \includegraphics[width=\textwidth]{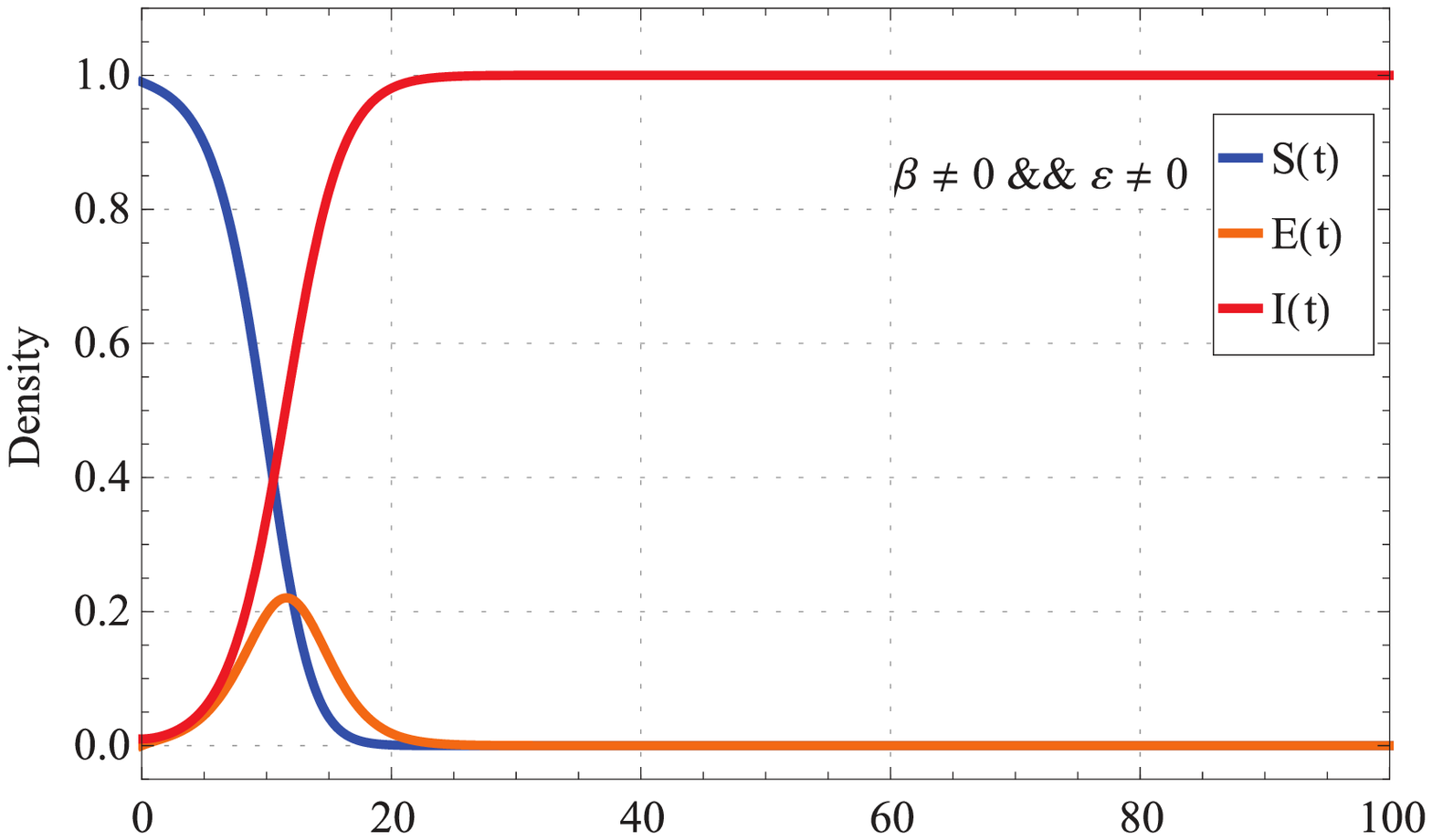}
                \caption{$S(0)=0.99$, $E(0)=0$, $I(0)=0.01$, $\beta=0.8$ and $\varepsilon=0.5$.}
                \label{Fig21b}
        \end{subfigure}%
        
        ~
        \begin{subfigure}[b]{0.485\textwidth}
                \centering
                \includegraphics[width=\textwidth]{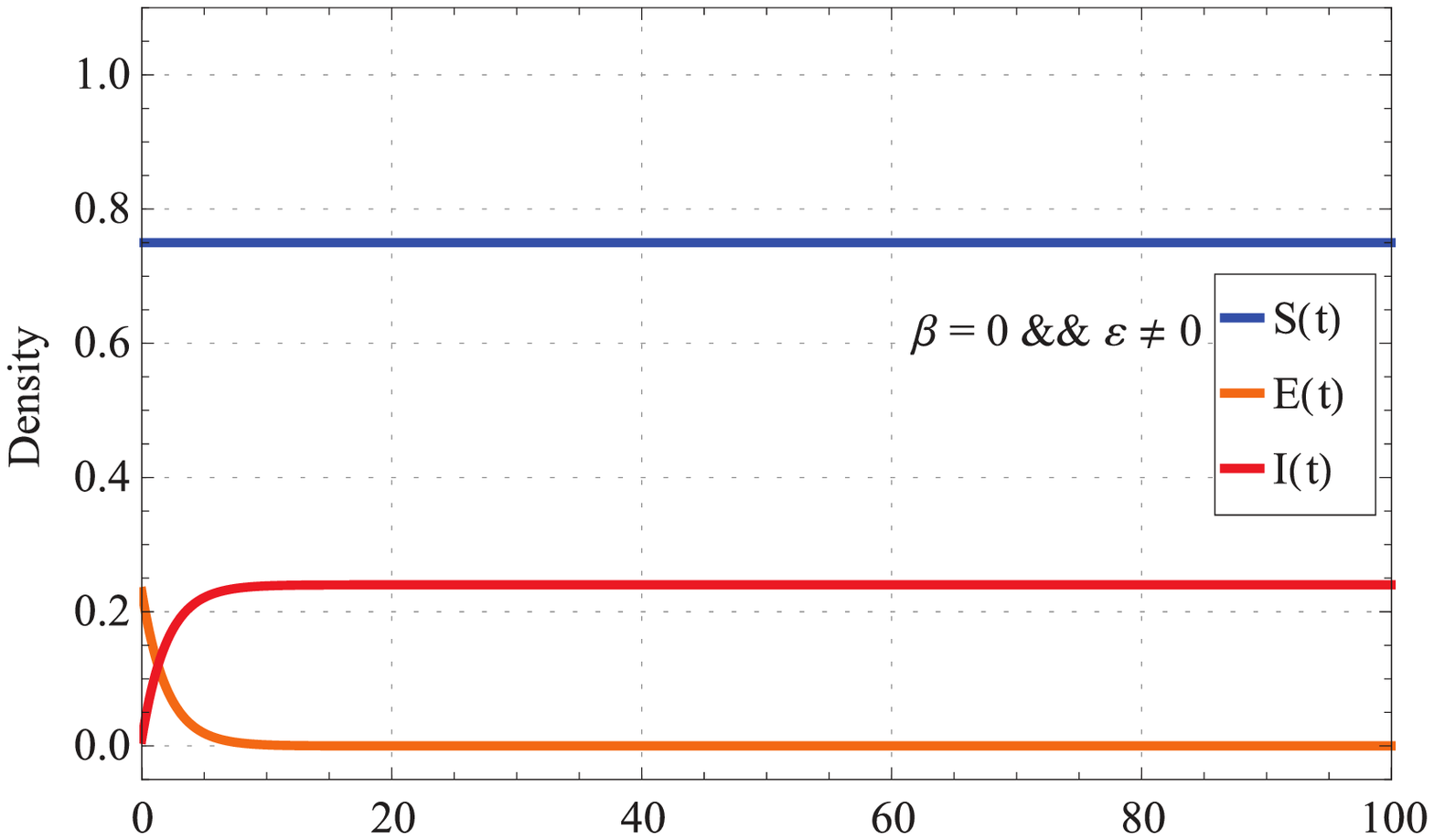}
                \caption{$S(0)=0.75$, $E(0)=0.24$, $I(0)=0.01$, $\beta=0$, and $\varepsilon=0.5$.}
                \label{Fig21c}
        \end{subfigure}%
        ~
        \begin{subfigure}[b]{0.485\textwidth}
                \centering
                \includegraphics[width=\textwidth]{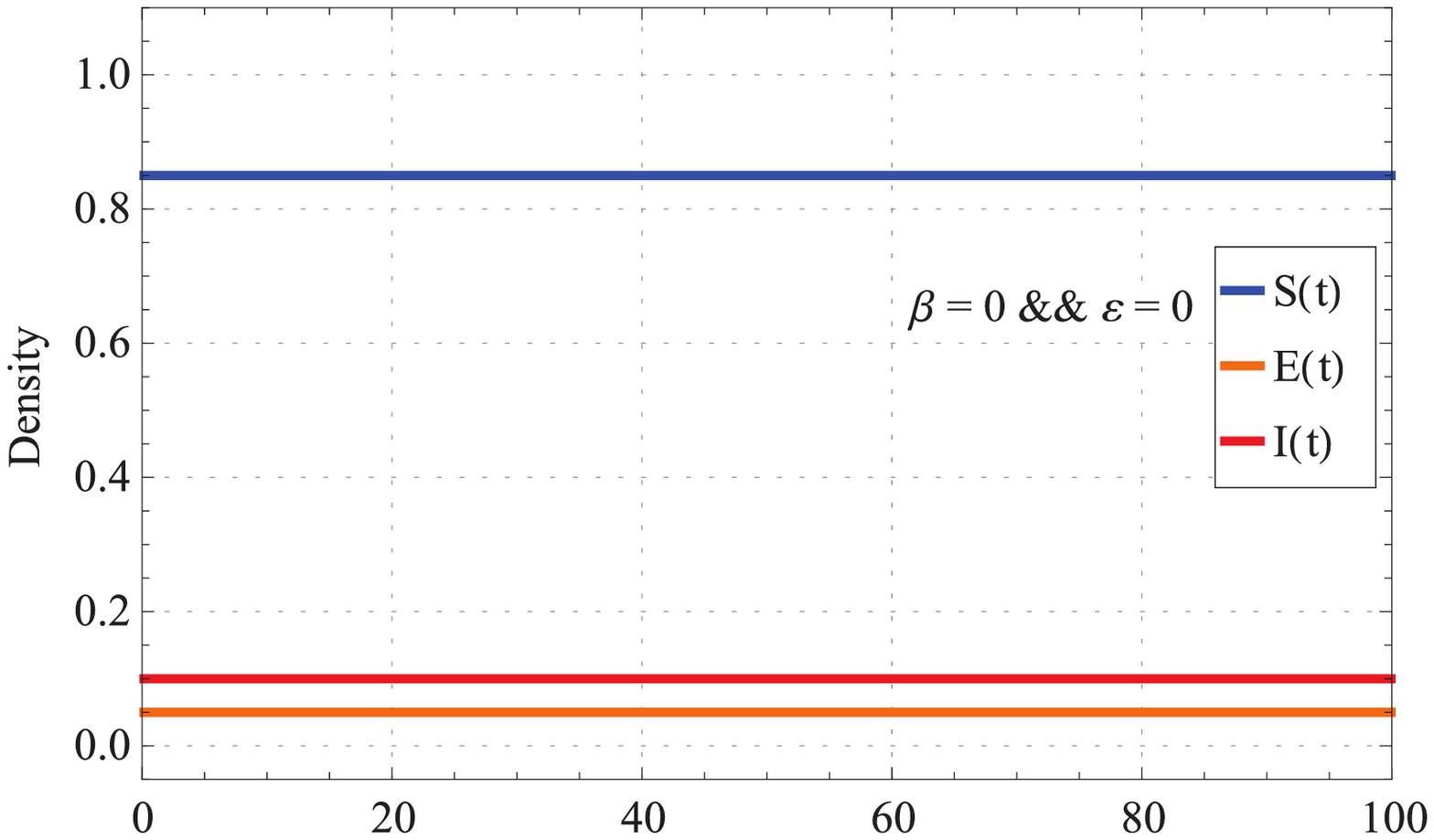}
                \caption{$S(0)=0.85$, $E(0)=0.05$, $I(0)=0.1$, $\beta=0$, and $\varepsilon=0$.}
                \label{Fig21d}
        \end{subfigure}%
        
        ~
        \begin{subfigure}[b]{0.485\textwidth}
                \centering
                \includegraphics[width=\textwidth]{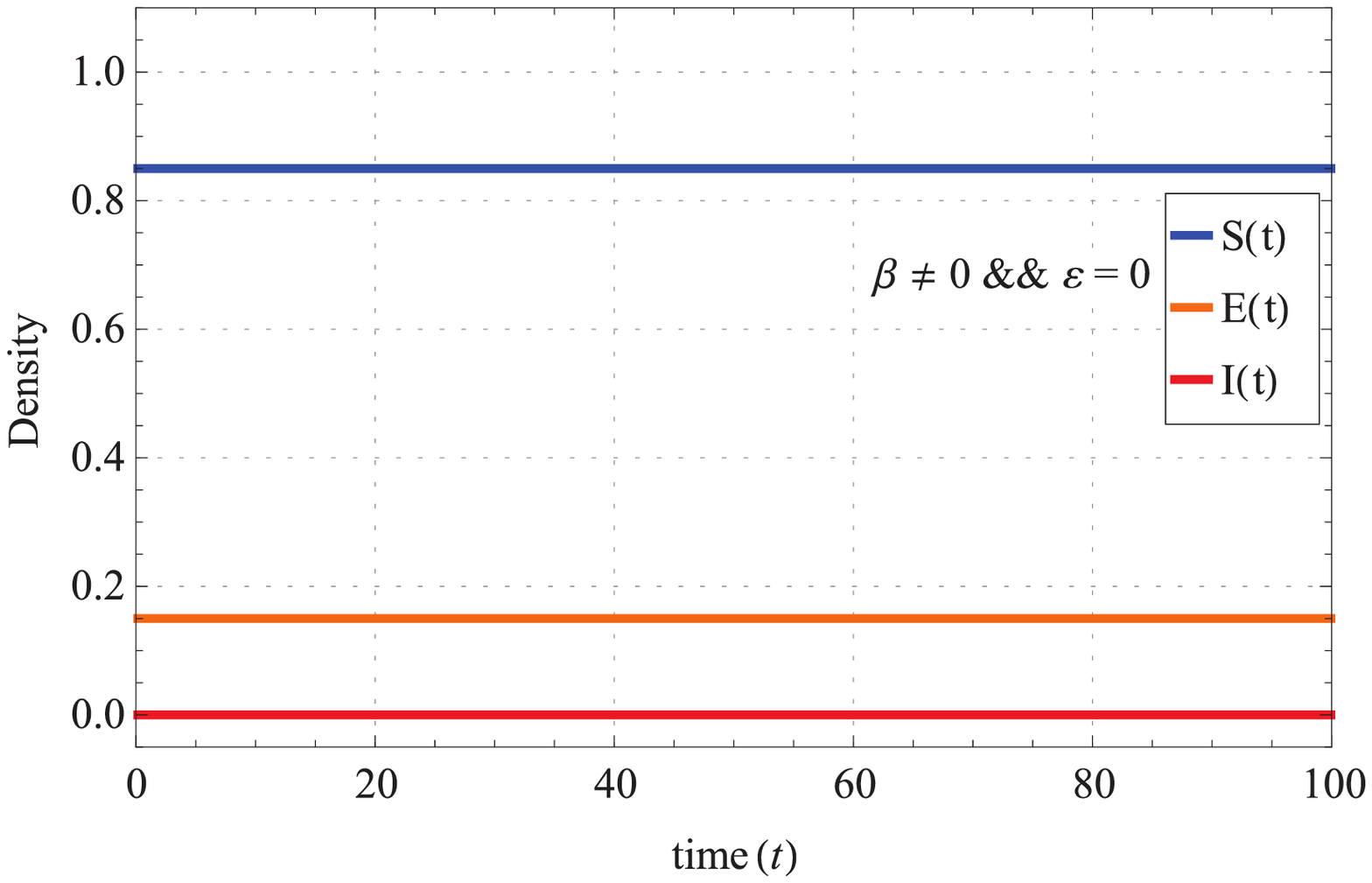}
                \caption{$S(0)=0.85$, $E(0)=0.15$, $I(0)=0$, $\beta=0.8$, and $\varepsilon=0$.}
                \label{Fig21e}
        \end{subfigure}%
        ~
        \begin{subfigure}[b]{0.485\textwidth}
                \centering
                \includegraphics[width=\textwidth]{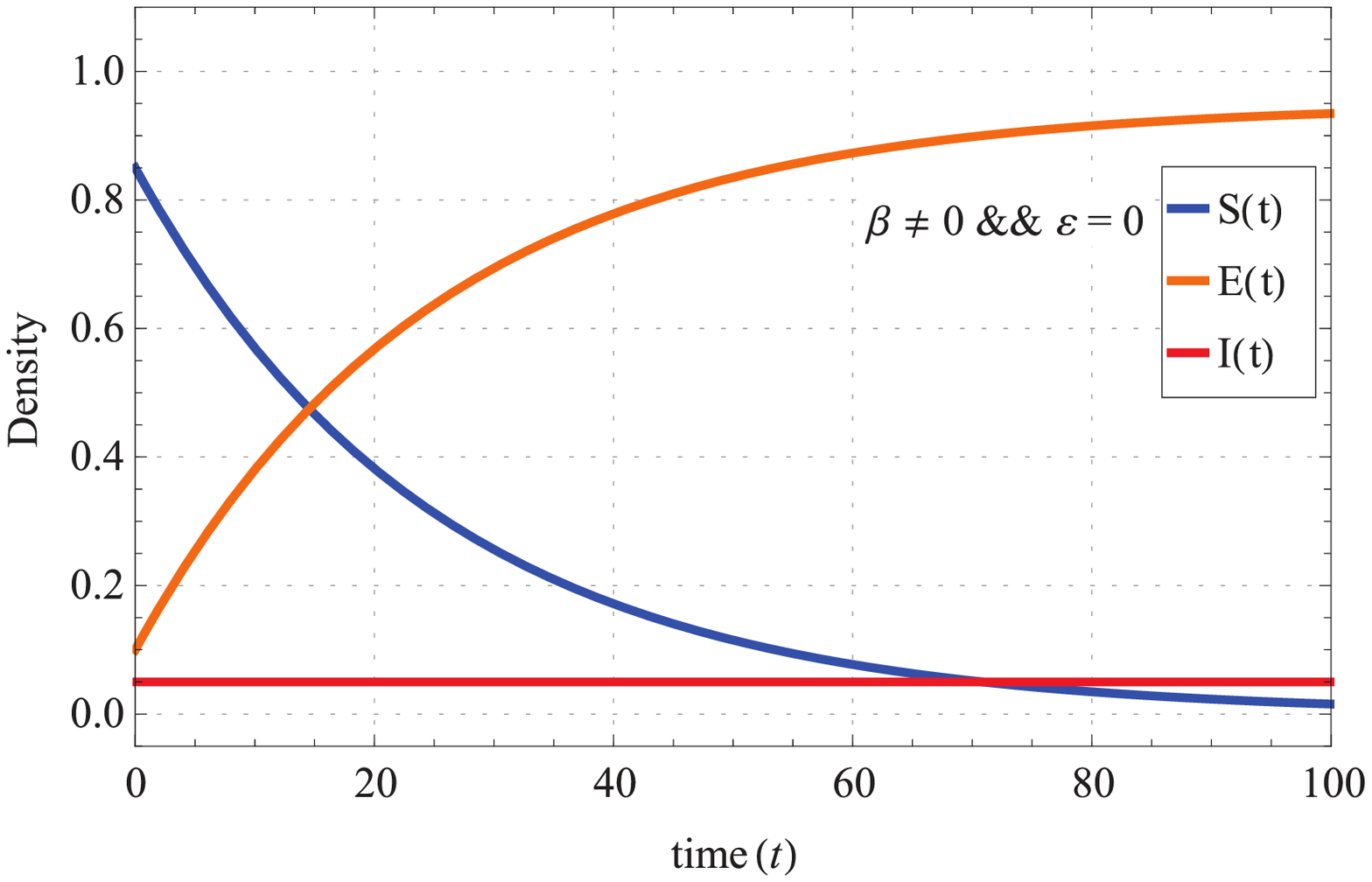}
                \caption{$S(0)=0.85$, $E(0)=0.1$, $I(0)=0.05$, $\beta=0.8$, and $\varepsilon=0$.}
                \label{Fig21f}
        \end{subfigure}%
        \caption[Density versus time for $SEI$ model without vital dynamics]{Density versus time for $SEI$ model without vital dynamics where (a) $E(0)=I(0)=0$ $(Case~A(i))$, (b) $E(0) \neq 0$ or $I(0) \neq 0$ $(Case~ A(ii))$, (c) $\beta=0$ and $\varepsilon \neq 0$ $(Case~B)$, (d) $\beta=\varepsilon=0$  $(Case~C)$, (e) $I(0)=0$ $(Case~D(i))$, and (f) $I(0)\neq0$ $(Case~D(ii))$.}
        \label{fig21_SEI_Densities}
\end{figure*}
		
	\item \emph{Case C:} When $\beta=0$ and $\varepsilon=0$, the solution is given as a set of constants. As $t$ approaches infinity, the system remains in the following equilibrium point as exemplified in Figure~\ref{Fig21d}, where $S(0)=0.85$, $E(0)=0.05$, and $I(0)=0.1$:
		\begin{eqnarray}
			\lim_{t \to \infty}(S(t),E(t),I(t))=(S(0),E(0),I(0)).			%(56)
		\end{eqnarray}
		
	\item \emph{Case D:} When $\beta \neq 0$ and $\varepsilon=0$, the solution of the reduced system of differential equations is given as below:
		\begin{equation}
		\label{eqn58_SEI_wvd}
  			\begin{split}
  			 	S(t)&= S(0),\\
				E(t)&= E(0)e^{-\varepsilon t},\\
				I(t)&= \varepsilon \int_0^t E(\tau) d\tau + I(0).			%(57)
  			\end{split}
		\end{equation}
		As Figures~\ref{Fig21e} and \ref{Fig21f} reveal, the equilibrium point that the system reaches depends upon the value of $I(0)$. Hence,
		\begin{enumerate}[(i)]
			\item If $I(0)=0$, we see that $S(t)=S(0)$ for all $t$. Thus,
					\begin{eqnarray}
						\lim_{t \to \infty}(S(t),E(t),I(t))&=(S(0),N-S(0),0).~~			%(58)
					\end{eqnarray}
			\item If $I(0)\neq 0$, $\lim_{t \to \infty}S(t)=0$ and we get:
					\begin{eqnarray}
						\lim_{t \to \infty}(S(t),E(t),I(t))&=(0,N-I(0),I(0)).~~			%(59)
					\end{eqnarray}
		\end{enumerate}
\end{enumerate}

\subsection{$SEI$ model with vital dynamics}
\begin{figure}[!t]
\centering
\includegraphics[width=2.3in]{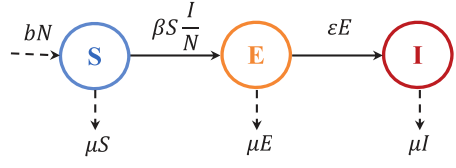}
\caption{$SEI$ model with vital dynamics.}
\label{fig22_SEI_vd}
\end{figure}
With $b=\mu$ in a birth-death population of total size $N$ , the model illustrated in Fig.~\ref{fig22_SEI_vd} can be written as follows:
\begin{eqnarray}
\label{eqn616263_SEI_vd}
\frac{dS}{dt}&{}={}& b N - \beta S \frac{I}{N} - \mu S,\\					%(60)
\frac{dE}{dt}&{}={}& \beta S \frac{I}{N} - \varepsilon E - \mu E,\\			%(61)
\frac{dI}{dt}&{}={}& \varepsilon E - \mu I.									%(62)
\end{eqnarray}

\subsubsection{Existence of equilibria}
\begin{figure*}[!t]
        \centering
        \begin{subfigure}[b]{0.485\textwidth}
                \centering
                \includegraphics[width=\textwidth]{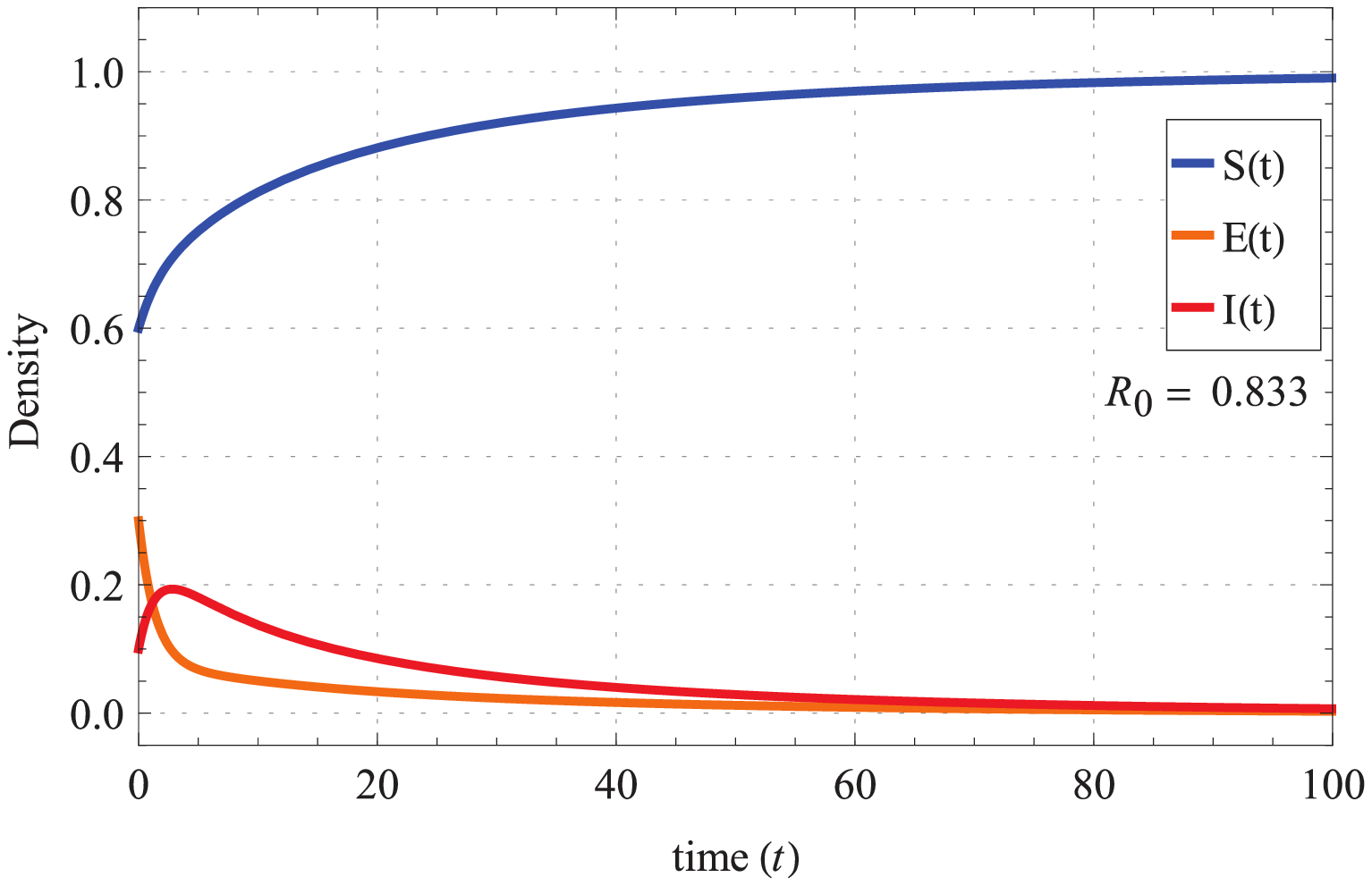}
                \caption{$\mathcal{R}_0=0.833$ with $\beta=0.25$, $\varepsilon=0.4$, and $\mu=0.2$.}
                \label{Fig23a}
        \end{subfigure}~ 
        ~  
        \begin{subfigure}[b]{0.485\textwidth}
                \centering
                \includegraphics[width=\textwidth]{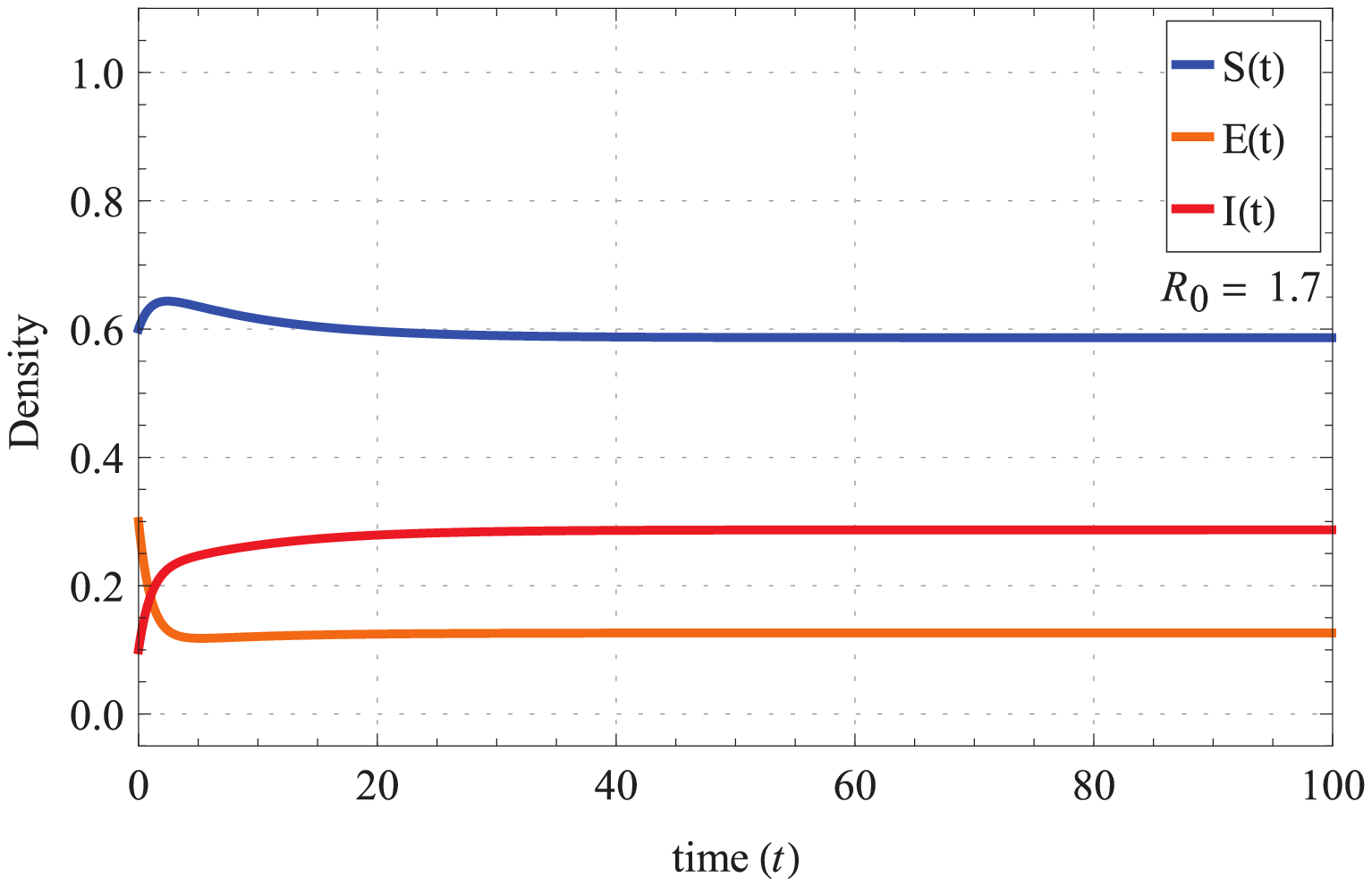}
                \caption{$\mathcal{R}_0=1.7$ with $\beta=0.54$, $\varepsilon=0.5$, and $\mu=0.22$.}
                \label{Fig23b}
        \end{subfigure}%
        \caption[Density versus time for $SEI$ model with vital dynamics]{Density versus time for $SEI$ model with vital dynamics where $N=1$, $S(0)=0.6$, $E(0)=0.3$, and $I(0)=0.1$.}
        \label{fig23_SEI_vd_Density}
\end{figure*}

The two set of equilibrium points obtained by setting the left-hand side of (60)-(62) to zero and solving for $S$, $E$, and $I$ are:
\begin{equation}
\label{eqn64_SEI_vd}
  \begin{split}
  	e_1:(S^*,E^*,I^*)&=(N, 0, 0),\\					%(63)
	e_2:(S^*,E^*,I^*)&=\left(\frac{N}{\mathcal{R}_0}, \frac{N}{\mathcal{R}_0} c_1(\mathcal{R}_0-1), \frac{N}{\mathcal{R}_0} c_2 (\mathcal{R}_0-1)\right),
  \end{split}
\end{equation}
with $c_1$ and $c_2$ defined as $\mu/(\varepsilon+\mu)$ and $\varepsilon/(\varepsilon+\mu)$, respectively, and $\mathcal{R}_0=\beta \varepsilon/(\mu(\varepsilon+\mu))$. The $DFE$ and $EE$ steady-states are respectively, $e_1$ and $e_2$. With $\beta=0.25$, $\varepsilon=0.4$, and $\mu=0.2$, Figure~\ref{Fig23a} depicts the disease-free steady state of the system as $\mathcal{R}_0=0.833$. Likewise, Figure~\ref{Fig23b} shows how the system reaches the endemic equilibrium for $\mathcal{R}_0=1.7$ when $\beta=0.54$, $\varepsilon=0.5$, and $\mu=0.22$.

\subsubsection{Equilibria stability analysis}
Considering (60) and (62), the Jacobian matrix for the system is as given below:
\renewcommand{\arraystretch}{2.0}
\begin{equation}
\label{eqn65_SEI_vd}
J=\begin{bmatrix}
		-\beta \dfrac{I}{N} - \mu &~~ -\beta \dfrac{S}{N}\\			%(64)
         -\varepsilon &~~ -(\varepsilon + \mu)
\end{bmatrix}.
\end{equation}

Evaluating the matrix at $e_1$ and solving the characteristic equation gives the following two eigenvalues:
\begin{eqnarray}
\lambda_{1,2}|_{e_1}=-\frac{1}{2}\left(\varepsilon+2\mu\pm\sqrt{\varepsilon(4\beta+\varepsilon)}\right). %(65)
\end{eqnarray}

For $e_1$ to be a stable node, both eigenvalues should be negative which means that $\lambda_2$ should be less than zero. Hence, $\varepsilon(4\beta+\varepsilon)$ should be less than $(\varepsilon+2\mu)^2$, which implies that $\mathcal{R}_0<1$. In other words, $e_1$ is a stable point if $\mathcal{R}_0<1$ and is a saddle point if the eigenvalues have opposite signs. Doing the same for $e_2$, we get:
\begin{eqnarray}
\lambda_{1,2}|_{e_2} = \frac{-\beta\varepsilon-(\varepsilon+\mu)^2}{2(\varepsilon+\mu)} \pm 
  \sqrt{\frac{(\beta\varepsilon)^2}{4(\varepsilon+\mu)^2} - \frac{\beta\varepsilon}{2} + \frac{(\varepsilon + \mu)(\varepsilon + 5\mu)}{4}}.
\end{eqnarray}

Similar to the above case, if $\mathcal{R}_0>1$, then both eigenvalues are negative and thus, $e_2$ would be a stable node. Otherwise, $e_2$ would be a saddle point. In the simplest case, with $\mathcal{R}_0$ as the bifurcation parameter, the system exhibits a forward transcritical bifurcation as it switches between the two equilibria. However, not much has been done in bifurcation analysis of such models in presence of other factors.

%~~~~~~~~~~~~~~~~~~~~~~~~~~~~~~~~~~~~~~~~~~~~~~~~~~~~~~~~~~~~~~~~~~~~~~~~~~~~~~~~~~~~~~~~~~~~~~~~~~~~~~~~~
%
%
%
%
%
%~~~~~~~~~~~~~~~~~~~~~~~~~~~~~~~~~~~~~~~~~~~~~~ SEIS Model ~~~~~~~~~~~~~~~~~~~~~~~~~~~~~~~~~~~~~~~~~~~~~~~
\section{The $SEIS$ model}
The \emph{susceptible}-\emph{exposed}-\emph{infected}-\emph{susceptible} ($SEIS$) model is an extension of the $SEI$ model such that in this model, the individual does not remain infected forever, but instead recovers and returns back to being susceptible again. Many sexually transmitted diseases (STD) and chlamydial infections are known to result in little or no acquired immunity following recovery \cite{Anderson1979}. In such cases, this model may serve as a suitable choice.
\begin{figure}[!t]
\centering
\includegraphics[width=2.0in]{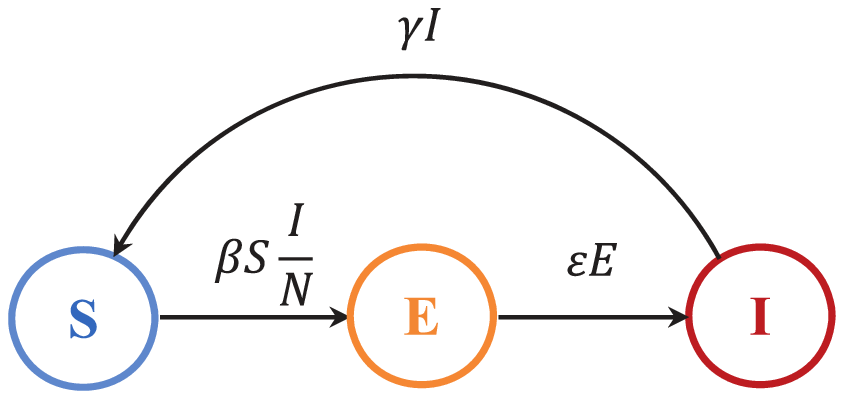}
\caption{The $SEIS$ model without vital dynamics.}
\label{fig24_SEIS_wvd}
\end{figure}

\subsection{$SEIS$ model without vital dynamics}
The dynamical transfer of hosts depicted in Figure~\ref{fig24_SEIS_wvd} can be formulated as follows, where $N=S+E+I$:
\begin{eqnarray}
\label{eqn686970_SEIS_wvd}
\frac{dS}{dt}&{}={}& \gamma I - \beta S \frac{I}{N},\\				%(67)
\frac{dE}{dt}&{}={}& \beta S \frac{I}{N} - \varepsilon E,\\			%(68)
\frac{dI}{dt}&{}={}& \varepsilon E - \gamma I.						%(69)
\end{eqnarray}

\subsubsection{Existence of equilibria}
\begin{figure}[!t]
        \centering
        \begin{subfigure}[b]{0.485\textwidth}
                \centering
                \includegraphics[width=\textwidth]{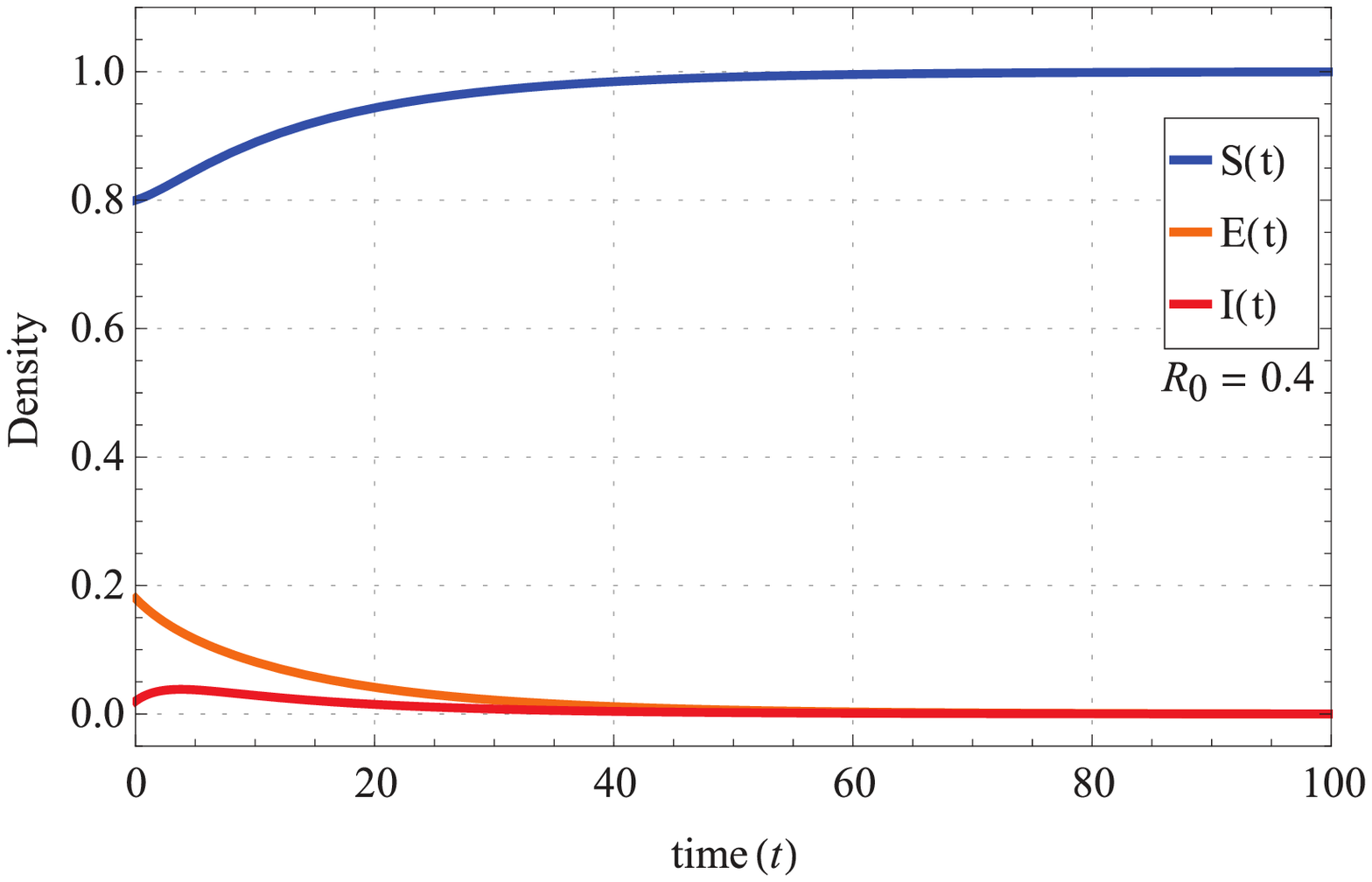}
                \caption{$\mathcal{R}_0=0.4$, $\beta=0.16$, $\varepsilon=0.12$, and $\gamma=04$.}
        \end{subfigure}~   
        ~
        \begin{subfigure}[b]{0.485\textwidth}
                \centering
                \includegraphics[width=\textwidth]{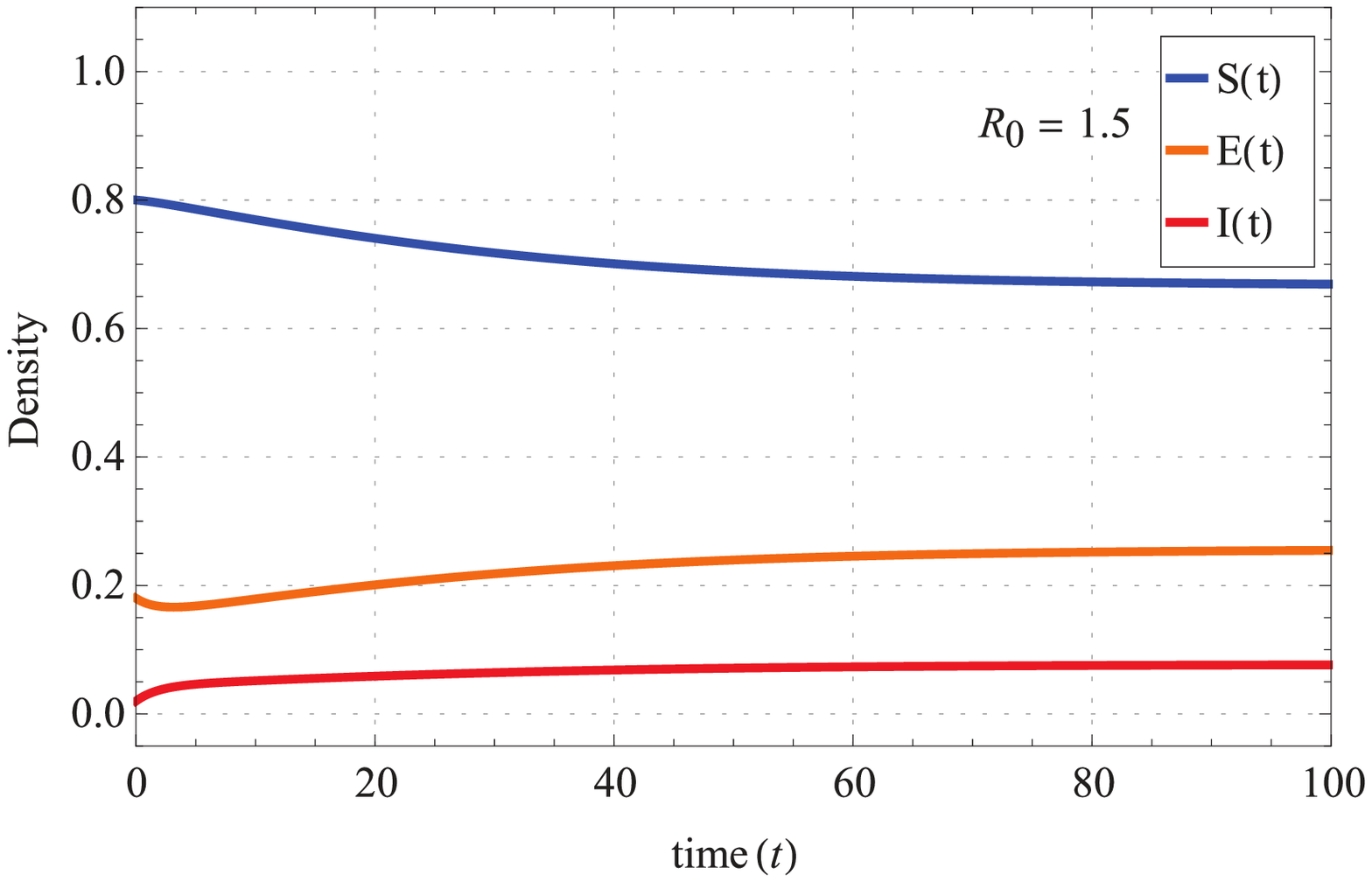}
                \caption{$\mathcal{R}_0=1.5$, $\beta=0.6$, $\varepsilon=0.12$, and $\gamma=0.4$.}
        \end{subfigure}%
        \caption[Density versus time for $SEIS$ model without vital dynamics]{Density versus time for $SEIS$ model without vital dynamics where $N=1$, $S(0)=0.8$, $E(0)=0.18$, and $I(0)=0.02$.}
        \label{fig25_SEIS_wvd_Density}
\end{figure}

On solving (67)-(69) for $S$, $E$, and $I$, we obtain $e_1$ and $e_2$ which represent the disease-free and endemic equilibrium points, respectively:
\begin{equation}
\label{eqn71_SEIS_wvd}
  \begin{split}
  	e_1:(S^*,E^*,I^*)&=(N, 0, 0),\\									%(70)
	e_2:(S^*,E^*,I^*)&=\left(\frac{N}{\mathcal{R}_0}, \frac{N}{\mathcal{R}_0} c_1 (\mathcal{R}_0-1), \frac{N}{\mathcal{R}_0} c_2 (\mathcal{R}_0-1)\right),						
  \end{split}
\end{equation}
where $c_1=\gamma/(\varepsilon+\gamma)$, $c_2=\varepsilon/(\varepsilon+\gamma)$, and $\mathcal{R}_0=\beta/\gamma$. As shown in Figure~\ref{fig25_SEIS_wvd_Density}, the system converges to $e_1$ when $\mathcal{R}_0 \leq 1$ and to $e_2$ when $\mathcal{R}_0>1$.

\subsubsection{Equilibria stability analysis}
\begin{figure}[!t]
        \centering
        \begin{subfigure}[b]{0.46\textwidth}
                \centering
                \includegraphics[width=\textwidth]{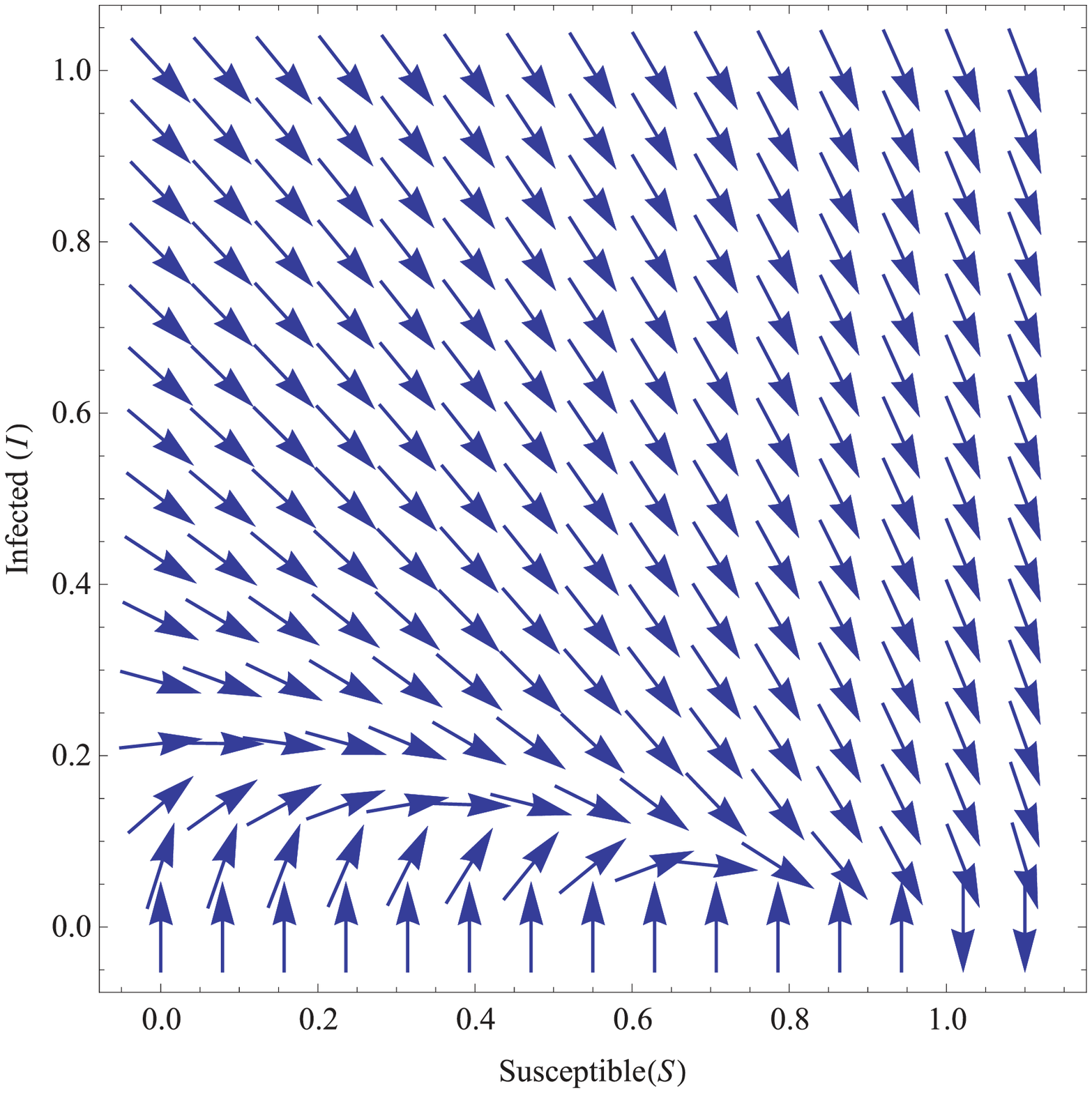}
                \caption{Stability of $e_1$ when $\mathcal{R}_0=0.4$, $\beta=0.16$, $\varepsilon=0.12$, and $\gamma=0.4$.}
                \label{Fig26a}
        \end{subfigure}~
        ~
        \vspace{0.1in}
        \begin{subfigure}[b]{0.46\textwidth}
                \centering
                \includegraphics[width=\textwidth]{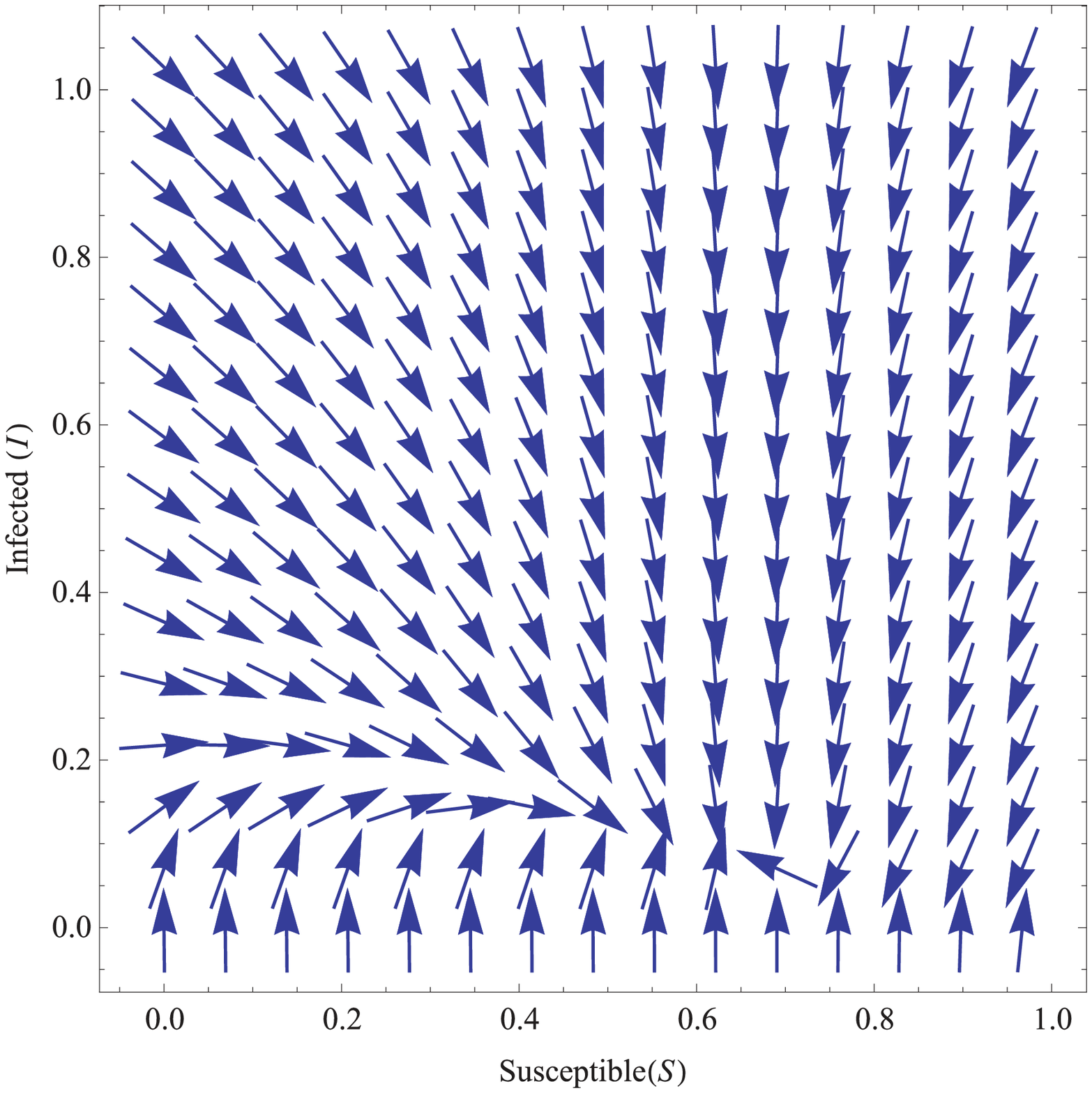}
                \caption{Stability of $e_2$ when $\mathcal{R}_0=1.5$, $\beta=0.6$, $\varepsilon=0.12$, and  $\gamma=0.4$.}
                \label{Fig26b}
        \end{subfigure}
        \caption[Stability of the equilibrium points in $SEIS$  model without vital dynamics]{Vector plots showing the stability of the equilibrium points in $SEIS$  model without vital dynamics for (a) $\mathcal{R}_0 \leq 1$ and ~(b) $\mathcal{R}_0 > 1$.}
        \label{fig26_SEIS_vector_wvd}
\end{figure}
The asymptotic behavior of the model can be analyzed by studying the stability conditions of the system near its equilibrium points. The Jacobian matrix formed by using (67) and (69) is as below:
\renewcommand{\arraystretch}{2.0}
\begin{equation}
\label{eqn72_SEIS_wvd}
J=\begin{bmatrix}
		-\beta \dfrac{I}{N} &~~ \gamma-\beta \dfrac{S}{N}\\			%(71)
         -\varepsilon &~~ -(\varepsilon + \gamma)
\end{bmatrix}.
\end{equation}

At $e_1$, the matrix $J$ yields the following eigenvalues:
\begin{eqnarray}
\lambda_{1,2}|_{e_1}=-\frac{1}{2}\left(\varepsilon+\gamma\pm\sqrt{(\gamma-\varepsilon)^2+4 \beta\varepsilon}\right).											%(72)
\end{eqnarray}

Once again, in order for $e_1$ to be stable, both eigenvalues should be negative and since $\lambda_1$ is already negative, $\lambda_2$ should be less than zero. Hence, for $\lambda_2$ to be negative, $-\gamma-\varepsilon + \sqrt{(\gamma-\varepsilon)^2+4\beta\varepsilon}$ should be negative, which on further simplification implies that $\mathcal{R}_0<1$. Therefore, $e_1$ is a stable $DFE$ when $\mathcal{R}_0<1$ and is unstable when $\mathcal{R}_0>1$. Similarly, calculating the eigenvalues of $J$ evaluated at $e_2$ results in the following pair of eigenvalues:
\begin{eqnarray}
\lambda_{1,2}|_{e_2}&{}={}&\left(-(\gamma+\varepsilon),~\frac{\gamma(\gamma-\beta)}		{(\gamma+\varepsilon)}\right),										%(73)
\end{eqnarray}

Since $\lambda_1$ is always negative, the stability of $e_2$ depends on $\lambda_2$. For $\lambda_2<0$, we observe that $e_2$ is a stable $EE$. On the contrary, when $\lambda_2>0$ (or equivalently, $\mathcal{R}>1$), the equilibrium point is unstable. This is clearly shown in Figure~\ref{fig26_SEIS_vector_wvd} where the stability of the system at the equilibrium points depends on $\mathcal{R}_0$. In Figure~\ref{Fig26a}, the system reaches the stable state $(S^*,I^*)=(1,0)$ for $\mathcal{R}_0=0.4$, whereas in Figure~\ref{Fig26b}, it converges to $(0.667,0.077)$ which is a stable endemic. As observed in previous cases, the model results in a forward bifurcation at $\mathcal{R}_0=1$ when switching from one steady-state to another. Thus, the stability of $e_1$ is persistent at $\mathcal{R}_0=1$.

\subsection{$SEIS$ model with vital dynamics}
\begin{figure}[!t]
\centering
\includegraphics[width=2.3in]{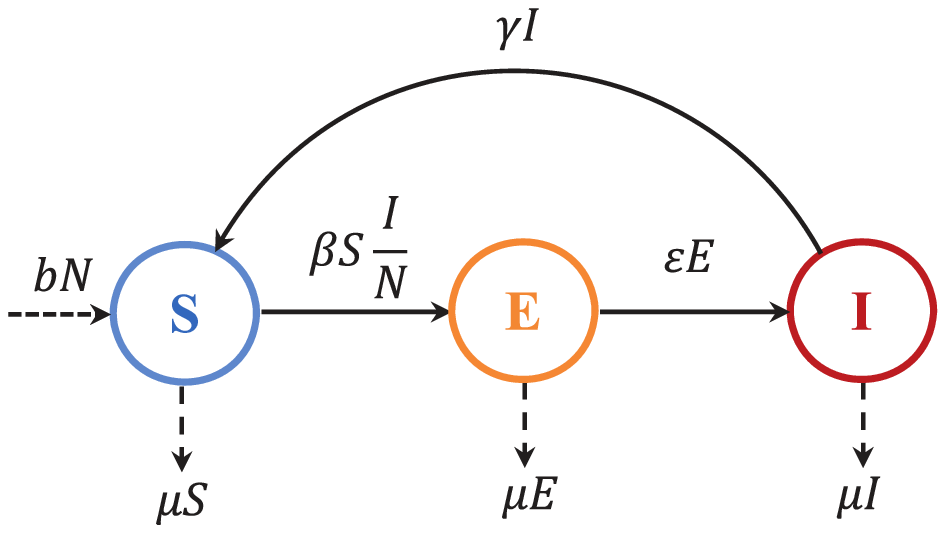}
\caption{The $SEIS$ model with vital dynamics.}
\label{fig27_SEIS_vd}
\end{figure}
In this subsection, we consider the $SEIS$ model for a population of size $N$ with birth and death rates that are constant. The set of equations given below refer to such a scheme illustrated in Figure~\ref{fig27_SEIS_vd}:
\begin{eqnarray}
\label{eqn757677_SEIS_wvd}
\frac{dS}{dt}&{}={}& b N +\gamma I - \beta S \frac{I}{N} - \mu S,\\			%(74)
\frac{dE}{dt}&{}={}& \beta S \frac{I}{N} - (\varepsilon+\mu) E,\\			%(75)
\frac{dI}{dt}&{}={}& \varepsilon E - (\gamma+\mu) I.						%(76)
\end{eqnarray}

\subsubsection{Existence of equilibria}
The following two equilibrium points are calculated by setting $b=\mu$, the time-derivatives in (74)-(76) to zero, and solving for $S$, $E$, and $I$:
\begin{equation}
\label{eqn78_SEIS_vd}
  \begin{split}
  	e_1:(S^*,E^*,I^*)&=(N, 0, 0),\\											%(77)
	e_2:(S^*,E^*,I^*)&=\left(\frac{N}{\mathcal{R}_0}, \frac{N}{\mathcal{R}_0} c_1 (\mathcal{R}_0-1), \frac{N}{\mathcal{R}_0} c_2 (\mathcal{R}_0-1)\right),
  \end{split}
\end{equation}
where $c_1$, $c_2$, and $\mathcal{R}_0$ are defined as $(\gamma+\mu)/(\gamma+\varepsilon+\mu)$, $\varepsilon/(\gamma+\varepsilon+\mu)$, and $\beta\varepsilon/((\varepsilon+\mu)(\gamma+\mu))$, respectively. For $\mathcal{R}_0 \leq 1$, the system approaches $e_1$ which is disease-free, while for $\mathcal{R}_0 > 1$, it ends up at $e_2$.

\subsubsection{Equilibria stability analysis}
The Jacobian matrix for this system is given as follows:
\renewcommand{\arraystretch}{2.0}
\begin{equation}
\label{eqn79_SEIS_vd}
J=\begin{bmatrix}
		-\beta \dfrac{I}{N} - \mu &~~ \gamma - \beta \dfrac{S}{N}\\			%(78)
        -\varepsilon &~~ -(\mu+\varepsilon+\gamma)
\end{bmatrix}.
\end{equation}

By evaluating the matrix in (78) at $e_1$, we see that the equilibrium point is a stable  disease-free equilibrium when both of the following eigenvalues are negative and is unstable if the eigenvalues have opposite signs. In terms of $\mathcal{R}_0$, $e_1$ is stable if $\mathcal{R}_0<1$ and unstable if $\mathcal{R}_0>1$:
\begin{eqnarray}
\lambda_{1,2}|_{e_1}=-\frac{1}{2}\left(\varepsilon+\gamma+2\mu\pm\sqrt{(\gamma-\varepsilon)^2+4 \beta\varepsilon}\right).													%(79)
\end{eqnarray}
Doing the same for $e_2$ yields a more complex pair of eigenvalues. However, on simplifying the eigenvalues, one can easily conclude that $e_2$ is stable when $\mathcal{R}_0>1$ and unstable otherwise. At $\mathcal{R}_0=1$, the system changes from $e_1$ to $e_2$ resulting in a forward bifurcation. A few works that reveal interesting bifurcation behaviors in $SEIS$ model can be found in \cite{Hethcote1991,Cao2011,Li2011}.

%~~~~~~~~~~~~~~~~~~~~~~~~~~~~~~~~~~~~~~~~~~~~~~~~~~~~~~~~~~~~~~~~~~~~~~~~~~~~~~~~~~~~~~~~~~~~~~~~~~~~~~~~~
%
%
%
%
%
%~~~~~~~~~~~~~~~~~~~~~~~~~~~~~~~~~~~~~~~~~~ Additional Factors ~~~~~~~~~~~~~~~~~~~~~~~~~~~~~~~~~~~~~~~~~~~
\section{Other factors in modeling}
Occurrences of certain events in nature and society may influence the behavior of the epidemic models seen so far. To guarantee that the model of choice mimics its counterpart in real-world, such influential factors should be taken into consideration. Many of these factors result in models with additional compartments which make them more complicated for mathematical analysis. In this section, a concise description on some of the most prominent factors that are considered in epidemic modeling is provided.

\subsection{Latent Period}
The time period between exposure and the onset of infectiousness is defined as the latent period. This is slightly different from the definition of \emph{incubation period} which is the time interval between exposure and appearance of the first symptom of the disease in question. Thus, the latent period could be shorter or even longer than the incubation period. As seen before, $SEI$ and $SEIS$ are two examples of models with latent period. Models with higher dimensions such as $SEIR$, $SEIRS$, $MSEIR$, and $MSEIRS$ have also been reported in the literature \cite{Ma2008}. However, since these models cannot be reduced to planar differential equation systems due to their complexity, only a few complete analytic results have been obtained.

\subsection{Quarantine and Vaccination}
In absence of vaccination for an outbreak of a new disease, isolation of diagnosed infectives and quarantine of people who are suspected of having been infected are some of the few control measures available. Models such as $SIQS$ and $SIQR$ with a quarantined compartment, denoted by $Q(t)$, assume that all infectives go through the quarantined compartment before recovering or becoming susceptible again \cite{Hethcote2002}. However, vaccination, if available, is one of the most cost-effective methods of preventing disease spread in a population. We refer the reader to \cite{Keeling2011}, \cite{Vynnycky2010}, and \cite{Arino2003} for more on models with vaccination and vaccine efficacy.

\subsection{Time Delay}
Models with time delay deal with the fact that the dynamic behavior of disease transmission at time $t$ depends not only on the current state but also on the state of previous time \cite{Ma2009}, \cite{Arino2006}. Time delays are of two types namely, \emph{discrete} (or \emph{fixed}) delay and \emph{continuous} (or \emph{distributed}) delay. In the case of discrete time delay, the behavior of the model at time $t$ depends on the state at time $t-\tau$ as well, where $\tau$ is some fixed constant. As an example, with $\tau$ being the latent period for some disease, the number of infectives at time $t$ also depends on the number of infectives at time $t-\tau$. On the contrary, the behavior of a model with continuous delay at time $t$ depends on the states during the whole period prior to $t$ as well.

\subsection{Age Structure}
Since individuals in different age groups have different infection and mortality rates, age is considered to be an important characteristic in modeling infectious diseases. Mostly, cases of sexually transmitted diseases (STDs) such as AIDS occur in younger individuals as they tend to be more active within or between populations. Likewise, malaria is responsible for nearly half of the death of infants under the age of 5 due to their weak immune system. Hence, such facts highlight the importance of age structure in epidemic modeling. Age-structured models are broadly classified into three types namely, \emph{discrete} \cite{Yicang2004}, \emph{continuous} \cite{Li2008}, and \emph{age groups} or \emph{stages} \cite{Hethcote1997}.

\subsection{Multiple Groups}
In epidemiology, multi-group models describe the spread of infectious diseases in heterogeneous populations where each heterogeneous host population can be divided into several homogeneous groups in terms of geographic distributions, models of transmissions, and contact patterns. One of the pioneer models with multiple groups was investigated by Lajmanovich \emph{et al.} in \cite{Lajmanovich1976} for the transmission of gonorrhea. However, recent studies include differentiation of susceptibility to infection (DS) due to genetic variation of susceptible individuals, variation in infectiousness, and disease spread in competing populations \cite{Ma2009}.

\subsection{Migration}
A central assumption in the classical models seen so far is that the rate of new infections is proportional to the mass action term. In these models, we assumed that the infected and susceptible individuals mix homogeneously. An increasingly important issue in epidemiology is how to extend these classical formulations to adequately describe the spatial heterogeneity in the distribution of susceptible and infected people and in the parameters of the spread of the infection observed in both, experimental data and computer simulations. Diffusion or migration of individuals in space are simply of two types: \emph{migration among different patches} and \emph{continuous diffusion in space}. In the former type, migration of individuals between patches depend on the connectivity of the patches. Models with migration between two patches \cite{Hethcote1976} and \emph{n} patches \cite{Wang2004} have been reported in the literature. The latter type, on the other hand, takes into account the fact that the distribution of individuals and their interactions depend not only on the time \emph{t}, but also the location in a given space.

\subsection{Non-linear Forces of Infection}
Most of the classical epidemic models admit threshold dynamics, i.e. a $DFE$ is stable if $\mathcal{R}_0<1$ and an $EE$ is stable if $\mathcal{R}_0>1$. However, Capasso \emph{et. al} \cite{Capasso1997} showed that it is likely possible for a $DFE$ and $EE$ to be stable simultaneously. Futhermore, periodic oscillations have been observed in the incidence of various diseases including mumps, chickenpox, influenza, and the like. The question that arises is why are classical epidemic models unable to capture these periodic phenomena? The main reason is the nature of the force of infection. Classical models frequently use mass incidence and standard incidence which imply that the contact rate and infection probability per contact are constant in time. Nonetheless, it is more realistic (with added complexity) to consider the force of infection as a periodic function in time. As a simple example, consider the $SIR$ model with a periodic incidence function $F(I,t)$, and birth and death rates taken to be $\mu$ as given below \cite{Diallo2008}:
\begin{eqnarray}
\label{nonlinearSIR}
\frac{dS}{dt}&{}={}& \mu N - F(I, t)\frac{S}{N} - \mu S,\\			%(80)
\frac{dI}{dt}&{}={}& F(I, t)\frac{S}{N} - (\gamma+\mu) I,\\			%(81)
\frac{dR}{dt}&{}={}& \gamma I - \mu R.								%(82)
\end{eqnarray}

In recent years, much attention has been given to the study and analysis of chaotic behavior in epidemic models with non-linear infection forces.

%~~~~~~~~~~~~~~~~~~~~~~~~~~~~~~~~~~~~~~~~~~~~~~~~~~~~~~~~~~~~~~~~~~~~~~~~~~~~~~~~~~~~~~~~~~~~~~~~~~~~~~~~~
%
%
%
%
%
%~~~~~~~~~~~~~~~~~~~~~~~~~~~~~~~~~~~~~~~~~~~~~~ Conclusion ~~~~~~~~~~~~~~~~~~~~~~~~~~~~~~~~~~~~~~~~~~~~~~~
\section{Conclusion}
Mathematical modeling of communicable diseases has received considerable attention over the last fifty years. A wide range of studies on epidemic models has been reported in the literature. However, there lacks a comprehensive study on understanding the dynamics of simple deterministic models through implementation. Aiming at filling such a gap, this work introduced some widely-appreciated epidemic models and studied each in terms of mathematical formulation, near equilibrium point stability analysis, and threshold dynamics with the aid of \emph{Mathematica}. In addition, important factors that may be considered for better modeling were also presented. We believe that this article would serve as a good starting point for readers new to this research area and/or with little mathematical background.

% or

% Use this command to get the appendices' numbers in "A", "B" instead of the
% default capitalized Roman numerals ("I", "II", etc.).
% However, the capital letter form may result in awkward subsection numbers
% (such as "A-A"). Capitalized Roman numerals are the default.
%\useRomanappendicesfalse
%
%%\appendices
%% you can choose not to have a title for an appendix
%% if you want by leaving the argument blank
%\section{}

\bibliographystyle{ieeetran}
\bibliography{Epidemic_Modelling_Tutorial}

\end{document}